\documentclass{vldb}
\pdfoutput=1
\usepackage{etex}

\usepackage[usenames,dvipsnames]{xcolor} 
\colorlet{color1}{NavyBlue}
\colorlet{color2}{Orange}
\colorlet{color3}{OliveGreen}
\colorlet{color4}{red}

\usepackage[T1]{fontenc}
\usepackage[scaled]{beramono}
\pdfmapfile{=operatorsymbols.map}
\usepackage{operatorsymbols}
\usepackage{helvet}				
\usepackage[eulergreek]{sansmath}	

\usepackage{amsmath}
\usepackage{amssymb}
\usepackage{amsfonts}
\usepackage{mathtools}
\usepackage{MnSymbol}
\usepackage{semantic}

\usepackage{tabularx}		
\usepackage{multirow}		
\usepackage{booktabs}		

\usepackage{listings}
\usepackage{hyperref}		
\usepackage{graphicx}		
\usepackage{subfigure}		
\usepackage{balance}		
\usepackage{siunitx}			

\usepackage{tikz}
\usepackage{tikz-qtree}
\usetikzlibrary{patterns}		
\usetikzlibrary{decorations}	

\usepackage{pgfplots}
\usepackage{pgfplotstable}	
\pgfplotsset {				
   compat=newest,
   tick label style = {font=\sansmath\sffamily\small},
   every axis label/.append style={font=\sffamily\small},
   legend style={font=\sffamily\small},
   label style={font=\sffamily\small},
   every node near coord/.append style={font=\sansmath\sffamily\small, fill=white, inner sep=2pt},
   grid style={gray!50}
}
\usepgfplotslibrary{units}		
\usepgfplotslibrary{groupplots}	
\tikzstyle{every node}=[font=\sffamily\small]

\begin{document}


\title{High-Speed Query Processing over High-Speed Networks}

\numberofauthors{4}

\author{
\alignauthor Wolf R\"odiger\\
	\affaddr{TU M\"unchen}\\
	\affaddr{Munich, Germany}\\
	\email{roediger@in.tum.de}
\alignauthor Tobias M\"uhlbauer\\
	\affaddr{TU M\"unchen}\\
	\affaddr{Munich, Germany}\\
	\email{muehlbau@in.tum.de}
\alignauthor Alfons Kemper\\
	\affaddr{TU M\"unchen}\\
	\affaddr{Munich, Germany}\\
	\email{kemper@in.tum.de}
\alignauthor Thomas Neumann\\
	\affaddr{TU M\"unchen}\\
	\affaddr{Munich, Germany}\\
	\email{neumann@in.tum.de}
}

\maketitle


\begin{abstract}
Modern database clusters entail two levels of networks: connecting CPUs and NUMA regions inside a single server in the small and multiple servers in the large. The huge performance gap between these two types of networks used to slow down distributed query processing to such an extent that a cluster of machines actually performed worse than a single many-core server. The increased main-memory capacity of the cluster remained the sole benefit of such a scale-out.

The economic viability of high-speed interconnects such as InfiniBand has narrowed this performance gap considerably. However, InfiniBand's higher network bandwidth alone does not improve query performance as expected when the distributed query engine is left unchanged. The scalability of distributed query processing is impaired by TCP overheads, switch contention due to uncoordinated communication, and load imbalances resulting from the inflexibility of the classic exchange operator model. This paper presents the blueprint for a distributed query engine that addresses these problems by considering both levels of networks holistically. It consists of two parts: First, \emph{hybrid parallelism} that distinguishes local and distributed parallelism for better scalability in both the number of cores as well as servers. Second, a \emph{novel communication multiplexer} tailored for analytical database workloads using remote direct memory access (RDMA) and low-latency network scheduling for high-speed communication with almost no CPU overhead. An extensive evaluation within the HyPer database system using the TPC-H benchmark shows that our holistic approach indeed enables \emph{high-speed query processing over high-speed networks}.
\end{abstract}


\section{Introduction}
\label{Section:Introduction}

Main-memory database systems have gained increasing interest in academia and industry over the last years. The success of academic projects, including MonetDB \cite{Manegold2009MonetDB} and HyPer \cite{Kemper2011HyPer}, has led to the development of commercial main-memory database systems such as Vectorwise, SAP HANA, Oracle Exalytics, IBM DB2 BLU, and Microsoft Apollo.

This development is driven by a significant change in the hardware landscape: Today's many-core servers often have main-memory capacities of several terabytes. The advent of these brawny servers enables unprecedented single-server query performance. Moreover, a small cluster of such servers is often already sufficient for companies to analyze their business. For example, Walmart---the world's largest company by revenue---uses a cluster of only 16~servers with 64~terabytes of main memory to analyze their business data \cite{Plattner2014TUM}.

\begin{figure}
\centering
\includegraphics[width=1.0\linewidth]{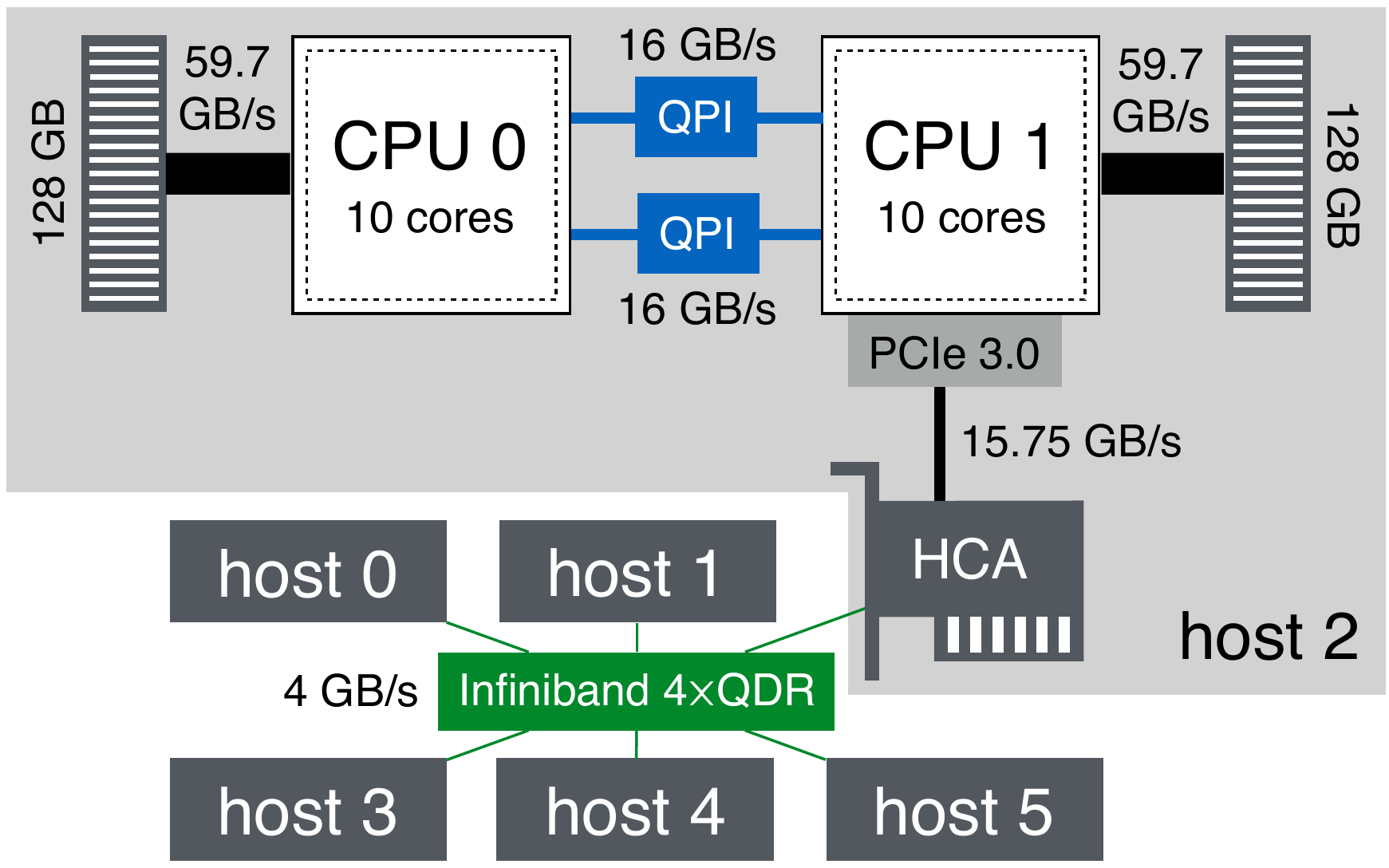}
\caption{Two levels of networks in a cluster: {\color{color1} connecting CPUs in the small} and {\color{color3} servers in the large}}
\label{Figure:Setup}
\end{figure}

Such a cluster entails two levels of networks as highlighted in Figure~\ref{Figure:Setup}: The network in the small connects several many-core CPUs and their local main memory inside a single server via a high-speed QPI interconnect. Main-memory database systems have to efficiently parallelize query execution across these many cores and adapt to the non-uniform memory architecture (NUMA) to avoid the high cost of remote memory accesses \cite{Li2013NUMAShuffling, Leis2014MorselDriven}. Traditionally, exchange operators are used to introduce parallelism both locally inside a single server as well as globally between servers. However, the inflexibility of the classic exchange operator model introduces several scalability problems. We propose a new \emph{hybrid approach} instead, that combines special decoupled exchange operators for distributed processing with the existing intra-server morsel-driven parallelism \cite{Leis2014MorselDriven} for local processing. Choosing the paradigm for each level that fits best, hybrid parallelism scales better with the number of cores per server than classic exchange operators as shown in Figure~\ref{Figure:Scalability}.

\begin{figure}
\centering
\begin{tikzpicture}
\begin{axis}[
	axis x line*=bottom,
	axis y line=left,
	height=5cm,
	width=8cm,
	xlabel={number of servers},
	ylabel={speed-up of\\query response times},
	ylabel style={align=center},
	xmin=1,
	xmax=6,
	ymin=0,
	ymax=3.8,
	ymajorgrids,
	ytick={0, 1, 2, 3, 4, 5},
	yticklabels={{0$\times$}, {1$\times$}, {2$\times$}, {3$\times$}, {4$\times$}, {5$\times$}},
	tick align=center,
	legend style={anchor=north west, draw=none, fill=none, at={(.02, 1.05)}, inner sep=0pt},
	legend cell align=left,
	legend columns=1
]
\addplot[ultra thick, mark=square*, color1] coordinates {
(1, 1.00)
(2, 1.55)
(3, 2.05)
(4, 2.52)
(5, 2.82)
(6, 3.5)
};
\addlegendentry{RDMA (40 Gb/s InfiniBand) + scheduling}
\addplot[ultra thick, mark=*, color3] coordinates {
(1, 1.00)
(2, 0.85)
(3, 1.06)
(4, 0.99)
(5, 1.20)
(6, 1.16)
};
\addlegendentry{TCP/IP (40\,Gb/s InfiniBand)}
\addplot[ultra thick, mark=triangle*, color2] coordinates {
(1, 1.00)
(2, 0.34)
(3, 0.20)
(4, 0.19)
(5, 0.18)
(6, 0.18)
};
\addlegendentry{TCP/IP (1\,Gb/s Ethernet)}
\end{axis}
\end{tikzpicture}
\caption{Simply increasing the network bandwidth is not enough; a novel RDMA-based communication multiplexer is required (HyPer, TPC-H, SF 100)}
\label{Figure:TPCHSpeedUp}
\end{figure}
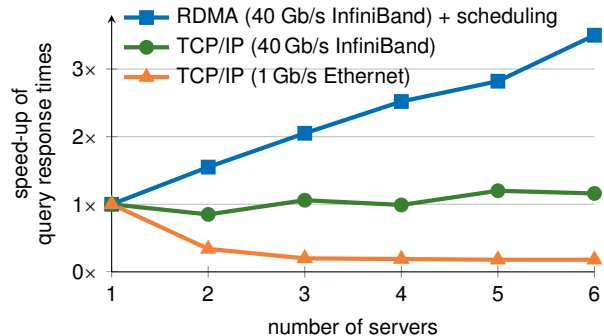

The network in the large connects separate servers. In the past, limited bandwidth actually \emph{reduced} query performance when scaling out to a cluster. Consequently, previous research focussed on techniques that avoid communication as much as possible \cite{Roediger2014NeoJoin, Polychroniou2014TrackJoin}. In the mean time, high-speed networks such as InfiniBand have become economically viable, offering link speeds of several gigabytes per second. However, faster networking hardware alone is not enough to scale query performance with the cluster size. Similar to the transition from disk to main memory, new bottlenecks surface when InfiniBand replaces Gigabit Ethernet. TCP/IP processing overheads and switch contention threaten the scalability of distributed query processing. Figure~\ref{Figure:TPCHSpeedUp} demonstrates these bottlenecks by comparing two distributed query engines using the TPC-H benchmark. Both engines are implemented in our in-memory database system HyPer. The first uses traditional TCP/IP, while the second is built with remote direct memory access (RDMA). The experiment adds servers to the cluster while keeping the data set size fixed at scale factor 100. Using Gigabit Ethernet actually decreases performance by 6$\times$ compared to using just a single server of the cluster. The insufficient network bandwidth slows down query processing. Still, a scale out is inevitable once the data exceeds the main memory capacity of a single server. InfiniBand 4$\times$QDR offers 32$\times$ the bandwidth of Gigabit Ethernet. However, Figure~\ref{Figure:TPCHSpeedUp} shows that simply using faster networking hardware is not enough. The distributed query engine has to be adapted to avoid TCP/IP overheads and switch contention. By combining RDMA and network scheduling in our novel distributed query engine we can scale query performance with the cluster size, achieving a speedup of $3.5\times$ for 6 servers. RDMA enables true zero-copy transfers at almost no CPU cost. Recent research has shown the benefits of RDMA for specific operators (e.g., joins \cite{Barthels2015RDMAJoin}) and key-value stores \cite{Kalia2014RDMAKeyValue}. However, we are the first to present the design and implementation of a complete distributed query engine based on RDMA that is able to process complex analytical workloads such as the TPC-H benchmark. In particular, this paper makes the following contributions:

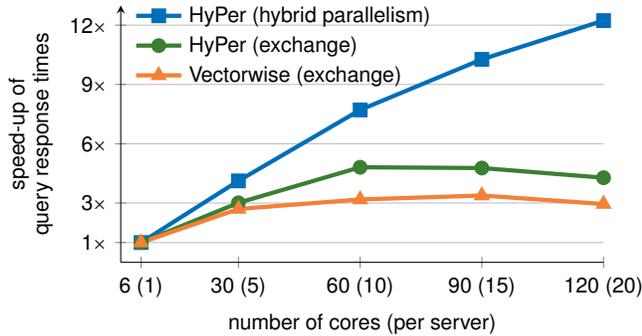
\begin{figure}[t]
\centering
\begin{tikzpicture}
\begin{axis}[
	height=5cm,
	width=8cm,
	axis x line*=bottom,
	axis y line=left,
	xlabel={number of cores (per server)},
	ylabel={speed-up of\\query response times},
	ylabel style={align=center},
	xmin=1,
	xmax=120,
	ymin=0,
	ymax=13,
	ymajorgrids,
	ytick={1, 3, 6, 9, 12},
	yticklabels={{1$\times$}, {3$\times$}, {6$\times$}, {9$\times$}, {12$\times$}},
	xtick={6, 30, 60, 90, 120, 150, 180, 210, 240},
	xticklabels={6 (1), 30 (5), 60 (10), 90 (15), 120 (20), 150, 180, 210, 240},
	tick align=center,
	legend style={anchor=north west, draw=none, fill=none, at={(.01, 1.04)}},
	legend cell align=left,
	legend columns=1
]
\addplot[ultra thick, mark=square*, color1] coordinates {
(6, 1.00)
(30, 4.12)
(60, 7.71)
(90, 10.27)
(120, 12.23)
};
\addlegendentry{HyPer (hybrid parallelism)}
\addplot[ultra thick, mark=*, color3] coordinates {
(6, 1.00)
(30, 3.01)
(60, 4.81)
(90, 4.77)
(120, 4.28)
};
\addlegendentry{HyPer (exchange)}
\addplot[ultra thick, mark=triangle*, color2] coordinates {
(6, 1.00)
(30, 2.70)
(60, 3.18)
(90, 3.38)
(120, 2.95)
};
\addlegendentry{Vectorwise (exchange)}
\end{axis}
\end{tikzpicture}
\caption{Hybrid parallelism scales significantly better with the number of cores per server than classic exchange operators (6 servers, TPC-H, SF 300)}
\label{Figure:Scalability}
\end{figure}

\begin{enumerate}
\item \emph{Hybrid parallelism:} A NUMA-aware distributed query execution engine that integrates seamlessly with intra-server morsel-driven parallelism, scaling considerably better both with the number of cores as well as servers compared to standard exchange operators.
\item A novel \emph{RDMA-based communication multiplexer} tailored for analytical database workloads that utilizes all the available bandwidth of high-speed interconnects with minimal CPU overhead; it avoids switch contention via low-latency network scheduling, improving all-to-all communication throughput by 40\,\%.
\item A prototypical implementation of our approach in our full-fledged in-memory DBMS HyPer that scales in both dimensions, the number of cores as well as servers.
\end{enumerate}

Section~\ref{Section:NetworkInTheLarge} evaluates high-speed cluster interconnects for typical analytical database workloads. Specifically, we study how to optimize TCP and RDMA for expensive all-to-all data shuffles common for distributed joins and aggregations. Building upon these findings, Section~\ref{Section:NetworkInTheSmall} presents a blueprint for our novel distributed query engine that is carefully tailored for both the network in the small and in the large. It consists of \emph{hybrid parallelism} for improved scalability in both the number of cores and servers as well as our \emph{optimized communication multiplexer} that combines RDMA and low-latency network scheduling for high-speed communication. Finally, Section~\ref{Section:Evaluation} provides a comprehensive performance evaluation using the ad-hoc OLAP benchmark TPC-H, comparing a prototypical implementation of our approach within our full-fledged in-memory database system HyPer to several SQL-on-Hadoop as well as in-memory MPP database systems: HyPer improves TPC-H performance by $256\times$ compared to Spark SQL, $168\times$ to Cloudera Impala, $38\times$ to MemSQL, and $5.4\times$ to Vectorwise Vortex.


\section{High-Speed Networks}
\label{Section:NetworkInTheLarge}

InfiniBand is a high-bandwidth and low-latency cluster interconnect. Several data rates have been introduced, which are compared to Gigabit Ethernet (GbE) in Table~\ref{Table:Interconnects}. The following performance study uses InfiniBand 4$\times$QDR hardware that offers 32$\times$ the bandwidth of GbE and latencies as low as 1.3 microseconds. We expect the findings to be valid for the faster data rates 4$\times$FDR and 4$\times$EDR as well.

InfiniBand offers the choice between two transport protocols: TCP via IP over InfiniBand (IPoIB) and the native InfiniBand \emph{ibverbs} interface for remote direct memory access (RDMA). In the following, we analyze and tune both protocols for analytical database workloads that require shuffling large amounts of data during distributed joins and aggregations. In contrast, transactional database workloads typically involve much smaller messages and would thus shift the tuning target from high throughput to low latencies.

\begin{table}
\centering
\newcolumntype{Y}{>{\raggedleft\arraybackslash}X}
\begin{tabularx}{\linewidth}{lrYYYYYY}
\toprule
& \multirow{2}{.7cm}[-.9mm]{\centering GbE} &  \multicolumn{5}{c}{InfiniBand (4$\times$)}\\
\cmidrule{3-7}
& & SDR & DDR & \textbf{QDR} & FDR & EDR\\
\midrule
{GB/s}			& 0.125	& 1		& 2		& \textbf{4}		& 6.8		& 12.1\\
{latency in $\mu$s}	& 340	& 5		& 2.5		& \textbf{1.3}		& 0.7		& 0.5\\
{introduction}		& 1998	& 2003	& 2005	& \textbf{2007}	& 2011	& 2014\\
\bottomrule
\end{tabularx}
\caption{Comparison of network data link standards}
\label{Table:Interconnects}
\end{table}


\subsection{TCP}

Existing applications that use TCP or UDP for network communication can run over InfiniBand using IPoIB. It is a convenient option to increase the network bandwidth for existing applications without changing their implementation.


\subsubsection{Data Direct I/O and Non-Uniform I/O Access}
\label{Section:DDIO}

Since the standardization of TCP in 1981 as RFC 793 the bandwidth provided by the networking hardware increased by several orders of magnitude. Yet, the socket interface still relies on the fact that message data is copied between application buffer and socket buffer \cite{Foong2003TCPPerformance}. The resulting multiple trips over the memory bus were identified as one of the main reasons hindering TCP scalability \cite{Clark1988TCPPerformance,Foong2003TCPPerformance,Frey2010iWARP}. However, we noticed during our experiments that this is no longer the case for modern systems. Indeed, the number of memory trips required by TCP and similar protocols was reduced significantly when Intel introduced data direct I/O (DDIO) in 2012 with its Sandy Bridge processors. DDIO allows the I/O subsystem to directly access the last level cache of the CPU for network transfers. DDIO has no hardware dependencies and is invisible to drivers and software.

\begin{figure}[t]
\centering
\subfigure[Classic I/O involves three memory trips at sender/receiver] {
	\centering
	\includegraphics[width=\linewidth]{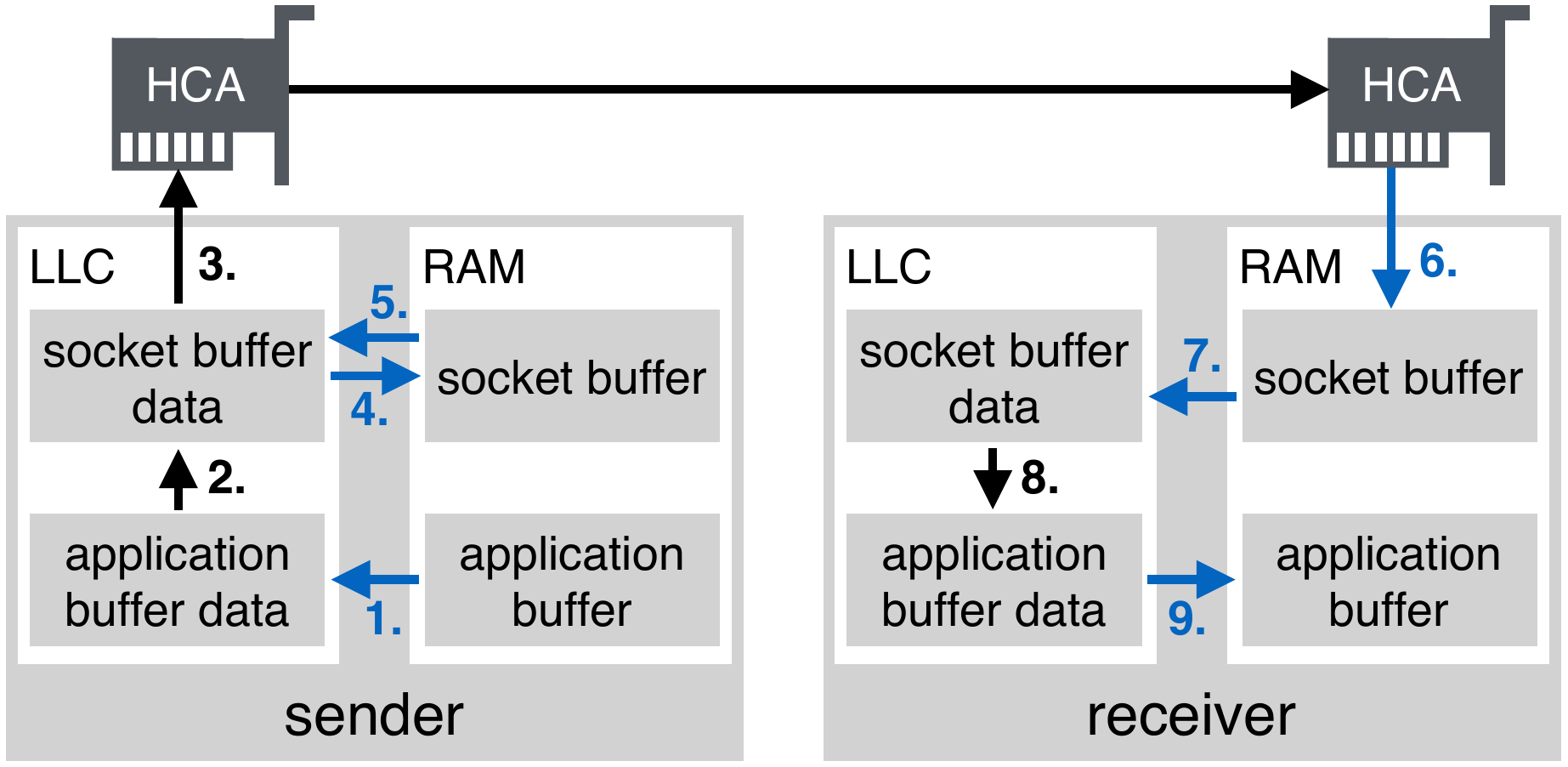}
	\label{Figure:ClassicIO}
}\\
\subfigure[Data direct I/O reduces this to only one memory trip each] {
	\centering
	\includegraphics[width=\linewidth]{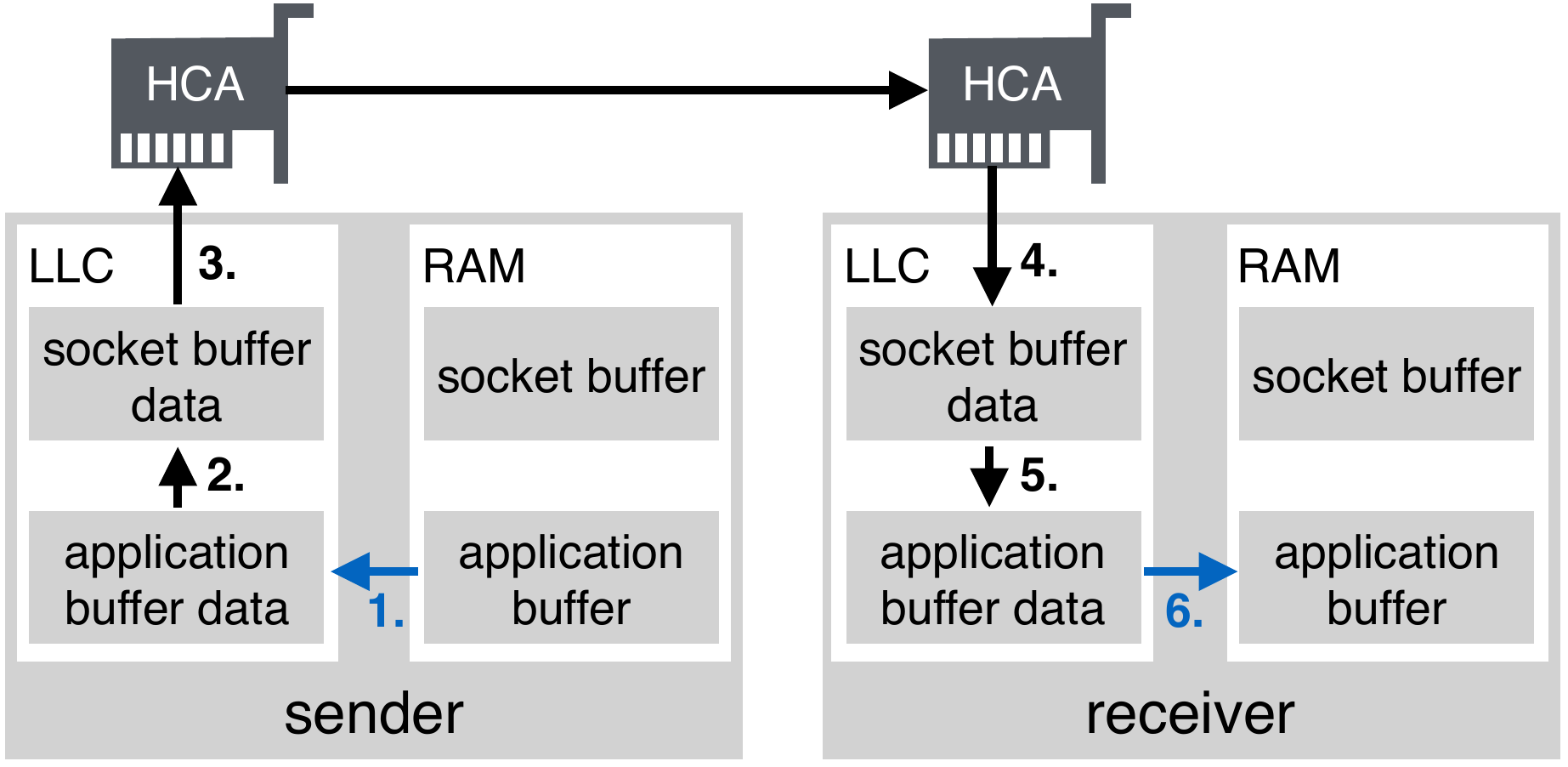}
	\label{Figure:DataDirectIO}
}
\caption{Data direct I/O significantly reduces the {\color{color1} memory bus traffic} for TCP compared to classic I/O}
\label{Figure:DDIO}
\end{figure}

Figure~\ref{Figure:ClassicIO} and \ref{Figure:DataDirectIO} show the memory trips performed by the classic I/O model and data direct I/O, respectively. At the sender, classic I/O (1)~reads the data from application buffer into the last level cache (LLC), (2)~copies it into the socket buffer, and (3)~sends it over the network forwarded from the LLC, which causes (4)~cache eviction and (5)~a speculative read. At the receiver, the data is (6)~DMAed to the socket buffer in RAM, (7)~copied into LLC, (8)~copied into the application buffer, and (9)~written to RAM.

DDIO instead targets the last level cache (LLC) of the CPU directly. At the sender, DDIO (1)~reads the application data, (2)~copies it into the socket buffer, and (3)~sends it directly from LLC. At the receiver, the data is (4)~allocated or overwritten in the LLC via Write Allocate/Update (restricted to 10\,\% of the LLC capacity to reduce cache pollution), (5)~copied into the application buffer, and (6)~written to main memory. DDIO reduces the number of memory bus transfers from 3 to 1 compared to the classic I/O model.

NUMA systems add a complication: A network adapter is directly connected to one of the CPUs. Consequently, there is a difference between local and remote I/O consumption. This is called Non-Uniform I/O Access (NUIOA). NUIOA is a direct consequence of the multi-CPU architecture of modern many-core servers. NUIOA systems restrict DDIO to threads running on the CPU local to the network card. We validated this by measuring the memory bus traffic in a micro-benchmark using Intel PCM\footnote{Intel PCM enables access to core and uncore performance monitoring units: \url{http://www.intel.com/software/pcm}}: Running the network thread on the local NUMA node caused every byte to be read 1.03$\times$ on the sender side and written 1.02$\times$ on the receiver side. Running the network thread on the remote NUMA node read every byte 2.11$\times$ for the sender while on the receiver side it was read 1.5$\times$ and written 2.33$\times$ (overheads might be due to TCP control traffic, retransmissions, cache invalidations, and prefetching \cite{Foong2003TCPPerformance}). This demonstrates that DDIO was only active for the NUMA-local thread running on the NUIOA-local CPU. Accordingly, our distributed query engine pins the network thread to the NUIOA-local CPU to avoid extra memory bus trips. This enables the I/O system to directly target the cache, further reducing the already lower memory bus traffic of RDMA.


\subsubsection{Tuning TCP for Analytical Workloads}

We designed a micro-benchmark that compares TCP with RDMA for database workloads. The message size determines the bottleneck for TCP performance. For small transfers, processing time is dominated by kernel overheads, sockets and protocol processing. For bulk transfers, data touching (i.e., checksums and copy) and interrupt processing account for most of the processing time \cite{Foong2003TCPPerformance}. Analytical query processing transfers large chunks of tuples during distributed joins and aggregations, we will thus focus on tuning TCP throughput for large packets, reducing the per-byte cost.

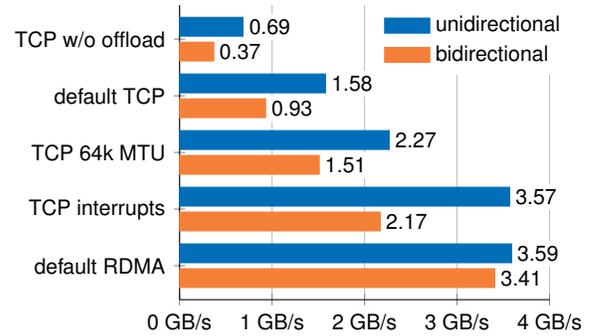
\begin{figure}[t]
\centering
\begin{tikzpicture}
\begin{axis}[
	xbar=0.08cm,
	bar width=0.25cm,
	enlarge y limits=0.15,
	axis x line*=bottom,
	axis y line*=left,
	width=6.5cm,
	height=5.5cm,
	xmin=0,
	xmax=4,
	xtick={0, 1, 2, 3, 4},
	xticklabels={0 GB/s, 1 GB/s, 2 GB/s, 3 GB/s, 4 GB/s},
	ytick=data,
	symbolic y coords={RDMA, INT, MTU, OFFLOAD, NONE},
	yticklabels={{default RDMA}, {TCP interrupts}, {TCP 64k MTU}, {default TCP}, {TCP w/o offload}},
	xmajorgrids,
	tick align=center,
	nodes near coords,
	nodes near coords align=horizontal,
	reverse legend,
	legend style={anchor=north west, draw=none, fill=white, at={(.53, 1)}},
	legend cell align=left,
	legend columns=1,
	area legend
]
\addplot[draw=color2, fill=color2] coordinates {
	(3.41,RDMA)
	(2.17,INT)
	(1.51,MTU)
	(0.93,OFFLOAD)
	(0.37,NONE)
};
\addlegendentry{bidirectional};
\addplot[draw=color1, fill=color1] coordinates {
	(3.59,RDMA)
	(3.57,INT)
	(2.27,MTU)
	(1.58,OFFLOAD)
	(0.69,NONE)
};
\addlegendentry{unidirectional};
\end{axis}
\end{tikzpicture}
\caption{Tuning TCP for analytical database workloads (one stream, 100k transfers, 512 KB messages)}
\label{Figure:TCP}
\end{figure}

Our micro-benchmark sends 100k distinct messages of size 512\,KB between two machines using a single thread. From a variety of TCP options that should improve performance only SACK gave a measurable improvement. SACK enables fast recovery from packet loss, which is especially relevant for high-speed links. In a first experiment we transfer data only from sender to receiver, while in the second we use fully duplex communication. The results are shown in Figure~\ref{Figure:TCP}.

The original specification of IPoIB in RFC 4391 and 4392 was limited to the datagram mode. This mode supports a 2,044 byte MTU, TCP offloading to the network card, and IP multicast. The connected mode was later added in RFC 4755. While it allows a MTU of up to 65,520 bytes, it does not support TCP offloading or IP multicasting. Disabling TCP offloading in datagram mode decreases the throughput for bidirectional transfer by 60\,\% from 0.93\,GB/s to 0.37\,GB/s. The connected mode with the same MTU of 2,044 bytes and without support for TCP offloading performs similar at 0.38\,GB/s, as one might expect. However, the larger MTU of 65,520 bytes available in connected mode more than offsets the missing TCP offloading features. The large MTU increases the throughput to 1.51\,GB/s, an improvement of 62\,\% over datagram mode with offloading.

A further performance bottleneck for TCP is interrupt handling. The network card issues interrupt requests (IRQ) to which the kernel responds by executing an interrupt handler on a configured core. The kernel automatically schedules the network thread to this same core to reduce cache misses. However, TCP throughput increases by a further 44\,\% to 2.17\,GB/s when the network thread is explicitly pinned to a different core. While this improves performance it also adds to the CPU overhead as now two cores are used.

We further investigated the impact of NUIOA on TCP throughput. In our micro-benchmark, pinning the network thread to the local socket improves throughput by 15\,\% in datagram and 6\,\% in connected mode for bidirectional transfers. The interrupt handler should always run on the same socket as the network thread or throughput drops by 50\,\%.

The bottleneck of TCP remains the CPU load of the receiver. Receive and send thread as well as the interrupt handler add a significant CPU overhead. The receiver experiences 100\,\%--190\,\% CPU utilization for the unidirectional transfers. The peak of 190\,\% (i.e., two fully occupied cores) is reached in datagram mode when network thread and interrupt handler are pinned to different cores.


\subsection{RDMA}

RDMA is InfiniBand's asynchronous, zero-copy communication method that incurs almost no CPU overhead and thereby frees resources for application-specific processing.


\subsubsection{Asynchronous Operation}

InfiniBand's \emph{ibverbs} interface is inherently asynchronous. Work requests are posted to send and receive work queues of the InfiniBand host channel adapter (HCA). The HCA processes work requests asynchronously and adds a work completion to a completion queue once it has finished. The application can process completion notifications when it sees fit to do so. This asynchronous interface makes overlapping of communication and computation easier than TCP, which would require two threads or the use of non-blocking sockets.

\subsubsection{Kernel Bypassing}

The InfiniBand HCA reads and writes main memory directly without interacting with the operating system or application during transfers. This avoids the overhead of system calls and the copying between application and system buffers. Consequently, the application has to manage buffers explicitly. For this purpose, RDMA introduces the concept of a memory region. A memory region provides the mapping between virtual and physical addresses so that the HCA can access main memory at any time without involving the kernel. Memory regions have to be registered beforehand to pin the memory and avoid swapping to disk. Registering memory regions is a time-consuming operation \cite{Frey2010iWARP} and regions should thus be reused whenever possible. Our distributed query engine implements this via a message pool.


\subsubsection{Memory vs. Channel Semantics}

RDMA allows to remotely read and write the main memory of a remote server without involving its CPU. These so-called \emph{memory semantics} requires that the initiator of the remote read/write has the memory key for the target memory region. A separate channel is required to exchange memory keys before communication can start. The alternative are \emph{channel semantics} via two-sided send and receive operations. The receiver posts receive work requests that specify the target memory region for the next incoming message. This eliminates the requirement to exchange memory keys before the transfer can start. There is no performance difference between one- and two-sided operations \cite{Frey2010iWARP}. An application should choose the semantics that fit best.

For our distributed query engine, it makes sense to use two-sided operations (channel semantics). First, two-sided operations do not require a separate communication channel to exchange memory keys. Second, the receiver is notified when new messages arrive and can process incoming tuples right away. One-sided operations (memory semantics) do not involve the receiver in the transfer at all. Making the receiver aware about incoming messages would thus require a separate communication channel or busy polling.


\subsubsection{Polling vs. Events for Completion Notifications}

RDMA with channel semantics provides two mechanisms to check for the availability of new messages. The first uses busy polling to check for new completion notifications. While this guarantees lowest latency it also occupies one core to 100\,\%. The second mechanism uses events to signal new completion notifications. The HCA raises an interrupt when a new message has arrived and wakes threads that are waiting for this event. The event-based handling of completion notifications reduces the CPU overhead to a mere 4\,\% for a full-speed bidirectional transfer with 512\,KB messages at the cost of a potentially higher latency compared to polling. Fortunately, the latency increase is insignificant for analytical database workloads with large messages.

\begin{figure*}[t]
\centering
\subfigure[] {
\centering
\begin{tikzpicture}[level distance=.8cm]
\Tree [.$\map$
	[.{$\group$}
		[.{$\join$}
			[.{$\join$}
				[.{part} ]
				[.{lineitem} ] ]
		        [.{\map}
			        [.{groupjoin}
					[.{part} ]
					[.{lineitem} ] ] ] ] ] ]
\end{tikzpicture}
\label{Figure:LocalPlan}
}
\hfil
\subfigure[] {
\centering
\begin{tikzpicture}[level distance=.7cm]
\Tree [.$\map$
	[.{$\group$}
		[.\textbf{\color{color1}exchange}
			[.{$\join$}
				[.{$\join$}
					[.\textbf{\color{color1}exchange}
						[.{part} ] ]
					[.\textbf{\color{color1}exchange}
						[.{lineitem} ] ]
				] 
			        [.{\map}
				        [.{groupjoin}
					        [.\textbf{\color{color1}exchange}
							[.{part} ] ]
						[.\textbf{\color{color1}exchange}
							[.{lineitem} ] ] ] ] ] ] ] ]
\end{tikzpicture}
\label{Figure:DistributedPlan}
}
\hfil
\subfigure[] {
\centering
\begin{tikzpicture}[level distance=.6cm]
\Tree [.$\map$
	[.{$\group$}
		[.\textbf{\color{color1}exchange}
			[.\textbf{\color{color3}pre-aggregation}
				[.{$\join$}
					[.\textbf{\color{color1}exchange}
						[.{$\join$}
							[.\textbf{\color{color2}broadcast}
								[.{part} ] ]
							[.{lineitem} ] ] ]
				        [.{\map}
					        [.{$\group$}
						        [.\textbf{\color{color1}exchange}
						        [.{groupjoin}
							        [.\textbf{\color{color2}broadcast}
									[.{part} ] ]
								[.{lineitem} ] ] ] ] ] ] ] ] ] ]
\end{tikzpicture}
\label{Figure:OptimizedPlan}
}
\caption{Query plans for TPC-H query 17: (a) for local execution, (b) for distributed execution introducing {\color{color1}exchange} operators where required, and (c) optimized with {\color{color3}pre-aggregation} and {\color{color2}broadcast} where beneficial}
\label{Figure:QueryPlans}
\end{figure*}
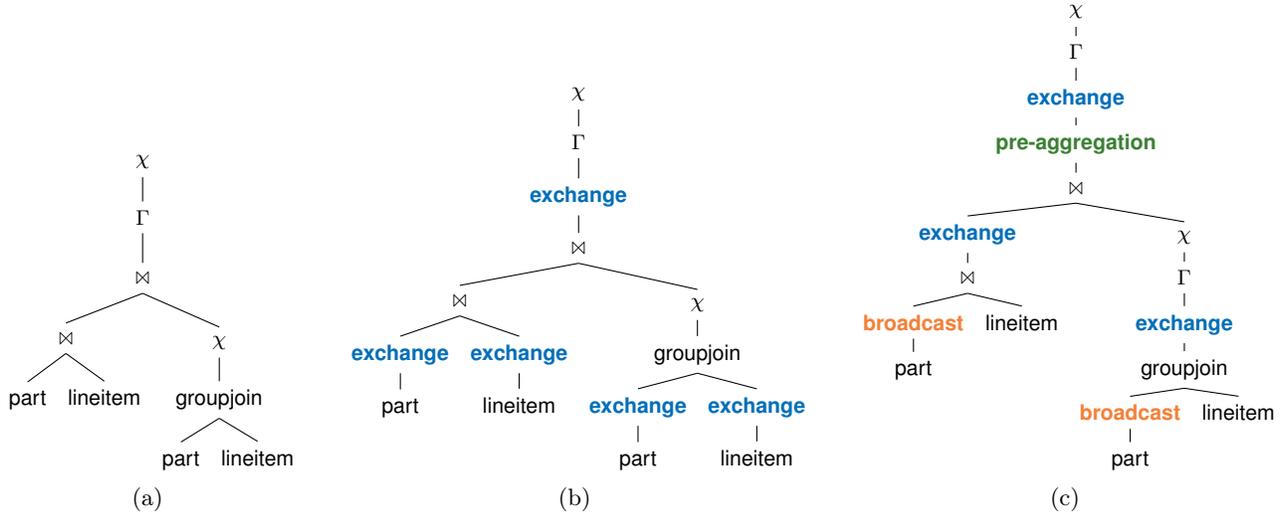


\subsection{Discussion}
\label{Section:NetworkDiscussion}

In the previous sections we discussed how to tune TCP and RDMA for analytical database workloads that shuffle large amounts of data between servers. This stands in contrast to transactional database workloads that typically involve much smaller messages and thus shift the focus from achieving maximal throughput to minimizing the latency.

While it is possible to bring TCP's throughput closer to that of RDMA via careful parameter tuning, this comes at the cost of significantly increased CPU load (100-190\,\% compared to 4\,\% for RDMA; single stream) and requires multiple streams to use all the available network bandwidth. RDMA is thus the better option as it enables truly asynchronous communication, requires less tuning, and frees the CPU for query processing. Our main findings for transmitting large messages over high-speed networks are the following:

\begin{enumerate}
\item \emph{Reduce memory traffic} by pinning the network thread to the NUIOA-local CPU, allowing the network card to target the cache directly and take advantage of DDIO.
\item \emph{For TCP:} Use IPoIB connected mode with the maximum MTU of 65,520 bytes, pin the network thread to a different core than the interrupt handler.
\item \emph{For RDMA:} Operate directly on message buffers for zero-copy communication, reuse buffers to avoid memory region registration costs, use channel semantics to simplify communication, use event-based completion notifications to minimize the CPU overhead.
\end{enumerate}


\section{High-Speed Query Processing}
\label{Section:NetworkInTheSmall}

The exchange operator is traditionally used to introduce parallelism both locally inside a single server as well as globally between the machines of a cluster. However, it introduces unnecessary materialization overheads for local processing, is inflexible when it comes to dealing with load imbalances, making it vulnerable to attribute value skew, and faces serious scalability issues due to the sheer number of parallel units, especially for modern many-core servers.

We propose a new hybrid approach instead, choosing the paradigm for each level that fits best. Locally, we use our existing morsel-driven parallelism \cite{Leis2014MorselDriven} to parallelize queries across cores ensuring NUMA-local processing. Globally, we designed a new data redistribution scheme between servers that combines decoupled exchange operators and a RDMA-based communication multiplexer that uses low-latency network scheduling. Both levels of parallelism are seamlessly integrated into a new hybrid approach that avoids unnecessary materialization, reacts to load imbalances at runtime, and scales better with the number of cores inside a single machine as well as the number of servers in the cluster.


\subsection{Classic Exchange Operators}

\begin{figure*}
\centering
\includegraphics[width=\linewidth]{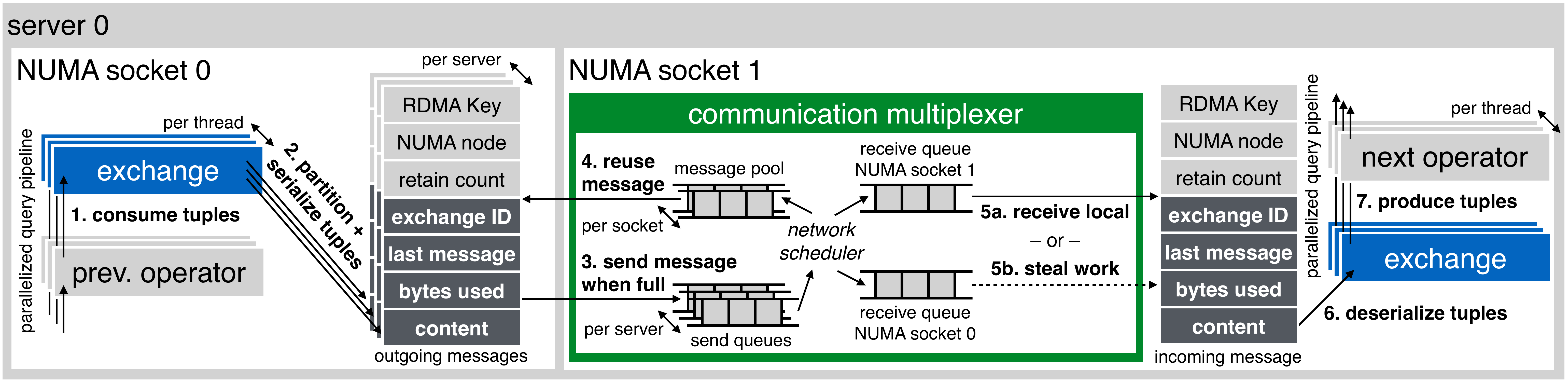}
\caption{Interaction of decoupled exchange operators with the RDMA-based, NUMA-aware multiplexer}
\label{Figure:HybridParallelism}
\end{figure*}

The exchange operator was introduced by Graefe for the Volcano database system \cite{Graefe1990Exchange}. It is a landmark idea as it allows systems to encapsulate parallelism inside an operator. All other relational operators are kept oblivious to parallel execution, making it straightforward to parallelize an existing non-parallel system. An example is shown in Figure~\ref{Figure:QueryPlans}: The single-server query plan shown in Figure~\ref{Figure:LocalPlan}, which was unnested by the query optimizer, is transformed into the distributed plan of Figure~\ref{Figure:DistributedPlan} by adding exchange operators where required. Two common optimizations for exchange operators are then introduced in Figure~\ref{Figure:OptimizedPlan}: First, instead of hash partitioning both inputs, the smaller input is broadcast when the inputs of a join have largely different sizes. Second, pre-aggregations significantly reduce the number of tuples that have to be shuffled---especially for aggregations with a small number of groups.

The exchange operator is commonly used to introduce parallelism both inside a single machine and between servers (e.g., Vectorwise Vortex \cite{Costea2012VectorwiseMPP} and Teradata \cite{Xu2008DataSkewJoin}). Threads execute copies of the query plan and are seen as separate parallel units that operate independently of each other. Parallel units communicate only via exchange operators. There is no difference between two parallel units that operate on the same server or on different machines. While this simplifies parallelization, it also introduces a number of problems.

The exchange operator fixes the degree of parallelism in the query plan, which makes it hard to deal with load imbalances and increases the impact of attribute value skew. Each exchange operator splits its input into one partition per parallel unit, e.g., 240 for our relatively small 6-server cluster with 20 hyper-threaded cores and thus 40 threads per machine. If one of the resulting 240~partitions contains more than $1/240$th of the input, all other parallel units have to wait for the straggler. A moderately skewed data set with Zipf factor $z = 0.84$ already more than doubles the input for the overloaded parallel unit. Hybrid parallelism instead distinguishes between local and remote parallel units and performs intra-server work stealing, reducing the number of parallel units to the number of servers. The same data set thus increases the input for the overloaded parallel unit by a mere 2.8\,\% for hybrid parallelism. The fewer parallel units, the lesser the impact of skew---orthogonal to the use of specific techniques that detect and deal with skew.

The large number of exchange operators in the classic approach also reduces the applicability of the broadcast optimization for distributed joins. A broadcast join is faster than hash partitioning when one input is much smaller than the other. This limit is $(n \times t) - 1$ for $n$ servers and $t$ local exchange operators per server as each exchange operator has to send its tuples to every other exchange operator. Hybrid parallelism distinguishes between local and distributed parallelism and can reduce this limit to $n - 1$ as the tuples need only be sent once to every remote server in the cluster. For our 6-server cluster, hybrid parallelism can thus use broadcast instead of hash joins already when the input sizes differ by 5$\times$ compared to 239$\times$ for classic exchange operators.

The huge number of connections and buffers required by the classic exchange operator model leads to further scalability issues and increases memory consumption significantly. An exchange operator requires a connection for each of the $n \times t - 1$ other exchange operators as well as a message buffer to partition its tuples. This results in $n^2 \times t^2 - t$ connections in the cluster and $n \times t - 1$ buffers per operator. For our relatively small 6-server cluster, this requires already a total of 57,560 connections in the cluster and 239 buffers per exchange operator. Hybrid parallelism instead integrates the exchange operators with intra-server morsel-driven parallelism and uses a dedicated communication multiplexer on each machine. It thereby eliminates the unnecessary materialization of intermediate results, significantly reduces the impact of skew, and requires only $n \times (n-1) = 30$ connections in the cluster and $n - 1 = 5$ buffers per exchange operator.


\subsection{Hybrid Parallelism}
\label{Section:HybridParallelism}

Figure~\ref{Figure:HybridParallelism} illustrates our new approach. It depicts the interaction of the decoupled exchange operators with the RDMA-based communication multiplexer during distributed query processing. Locally, query execution is parallelized according to our existing morsel-driven parallelism approach \cite{Leis2014MorselDriven} using one worker thread per hardware context of the server (two per core for hyper-threading). The input data---coming from either a pipeline breaker or a base relation---is split into work units of constant size, called morsels. Each worker pushes the tuples of its morsel all the way through the compiled query pipeline \cite{Neumann2011QueryCompilation} until a pipeline breaker is reached. This keeps tuples in registers and low-level caches for as long as possible. All database operators---including our new decoupled exchange operators---are designed such that workers can process the same pipeline job in parallel.

In detail, a decoupled exchange operator (1) consumes the tuples that are pushed to it by the preceding operator of its pipeline. It (2) partitions these tuples according to the CRC32 hash value of the join attributes into $n$ messages, one for each of the $n$ servers in the cluster. Broadcast exchange operators differ in that they instead serialize the tuples into a single message, using a retain counter to avoid multiple copies. A message consists of two parts: The first part includes its RDMA memory key, the NUMA node where the message resides, and said retain counter. Only the second part of a message is sent over the network: It consists of an identifier for the corresponding logical exchange operator, an indicator whether this is the last message for this operator, the number of bytes used, and the actual serialized tuples. Once a message is full or the exchange operator has processed all of its input, the message is (3) passed to the communication multiplexer and queued for sending. The exchange operator needs a new empty message that it (4) reuses from a memory pool, ensuring that it is NUMA-local to the CPU core on which the worker thread executes. The RDMA multiplexer sends and receives messages according to a round-robin network schedule to avoid link sharing and the resulting reduced network throughput. Only the used part of a partially-filled message is sent over the network. Once a message was successfully sent, it is put into the correct message pool for reuse. The multiplexer receives messages for every NUMA region in turn and notifies waiting exchange operators. These (5a) process NUMA-local messages. Only when there are none available, do they (5b) steal work from other NUMA regions. After (6) deserialization, the tuples are at last (7) pushed to the next operator in the pipeline.


\subsubsection{Decoupled Exchange Operators}

In contrast to the classic model, our decoupled exchange operator is unaware of all other exchange operators whether local or remote and only interacts with its communication multiplexer. This has several advantages: Our multiplexer sends broadcast messages once to every remote server. In contrast, classic exchange operators have to send it to every other exchange operator, reducing the applicability of broadcasts. The classic exchange operator is also inflexible in dealing with load imbalances as each operator is considered a separate parallel unit. Thus, skew has a much higher impact. Our hybrid approach instead treats servers as parallel units and uses work stealing inside the servers to handle load imbalances. Further, in the classic model each exchange operator needs a buffer for every other exchange operator compared to only one buffer per server for hybrid parallelism, which reduces memory usage significantly.

\begin{figure}
\centering
\includegraphics[width=\linewidth]{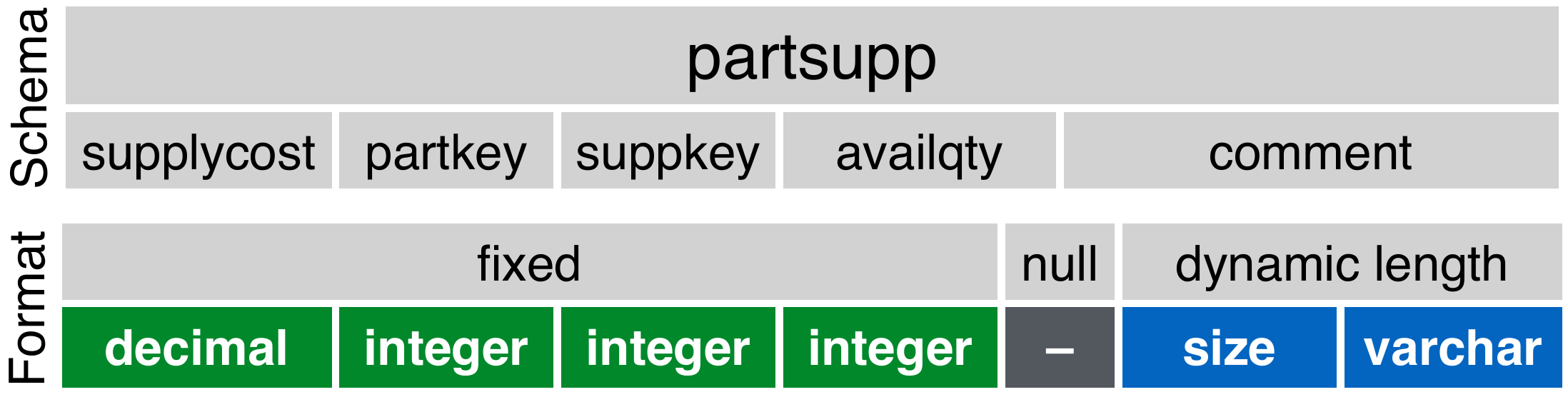}
\caption{Our densely-packed serialization format for the \emph{partsupp} relation of the TPC-H benchmark}
\label{Figure:Serialization}
\end{figure}

Our decoupled exchange operator uses LLVM code generation to efficiently serialize and deserialize tuples, minimizing the overhead of materialization. The code is expressly generated for the specific schema of the input tuples and thus does not need to dynamically interpret a schema. This reduces branching, improving code and data locality. Columns that are not required by subsequent operators are pruned as early as possible to reduce network transfer size. An example for our densely-packed, binary serialization format is shown in Figure~\ref{Figure:Serialization} for the \emph{partsupp} relation of the TPC-H benchmark. The format has three parts: The first part contains the values for all fixed-size attributes (e.g., decimal, integer, date) that are defined as \emph{not null} in a deterministic order determined first by the data type and second by the order in the schema. The second part consists of null indicators followed by the attribute values in case the attribute is not null for the current tuple. The third part contains the values for attributes of dynamic length (e.g., varchar, blob, text), which are stored as size and data content.


\subsubsection{RDMA-based, NUMA-aware Multiplexer}

Our novel communication multiplexer connects the decoupled exchange operators for distributed query processing. It uses RDMA and low-latency network scheduling for high-speed communication and ensures NUMA-local processing.

The multiplexer is a dedicated network thread per server that performs the data transfer between local and remote exchange operators by continuously sending messages according to a global round-robin schedule.  Local workers are not connected to all remote workers as this would lead to an excessive number of connections. Instead, only the multiplexers are connected with each other. Any available worker can process any incoming message. This enables work stealing and greatly alleviates the effect of skew. The multiplexer manages the send and receive queues as well as the reuse of messages via reference counting. Instead of deallocating messages when they are no longer needed, they are placed in a message pool. This avoids repeated memory allocation and deallocation during query processing as well as the expensive pinning of new messages to memory and registering them with the InfiniBand HCA to enable RDMA \cite{Frey2010iWARP}.

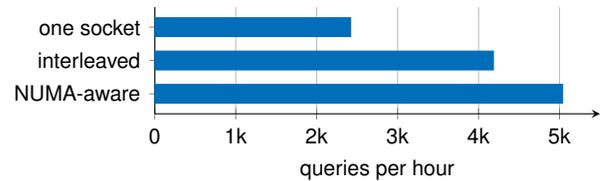
\begin{figure}
\centering
\begin{tikzpicture}
\begin{axis}[
	xbar=.2cm,
	bar width=0.25cm,
	enlarge y limits=0.3,
	height=3cm,
	width=7.5cm,
	axis y line*=left,
	axis x line=bottom,
	xlabel={queries per hour},
	symbolic y coords={AWARE, INTER, ONE},
	ytick=data,
	yticklabels={NUMA-aware, interleaved, one socket},
	xmin=0,
	xmax=5.5,
	xtick={0, 1, 2, 3, 4, 5},
	xticklabels={0, 1k, 2k, 3k, 4k, 5k},
	xmajorgrids,
	tick align=center
]
\addplot[fill=color1, draw=color1] coordinates {
(5.0392,AWARE)
(4.1831,INTER)
(2.4201,ONE)
};
\end{axis}
\end{tikzpicture}
\caption{Impact of NUMA-aware message allocation for a 4-socket server (HyPer, TPC-H, SF 100)}
\label{Figure:NUMA}
\end{figure}

\begin{figure*}

\subfigure[Round-robin scheduling with conflict-free phases; three phases for four servers] {
	\centering
	\includegraphics[width=.32\linewidth]{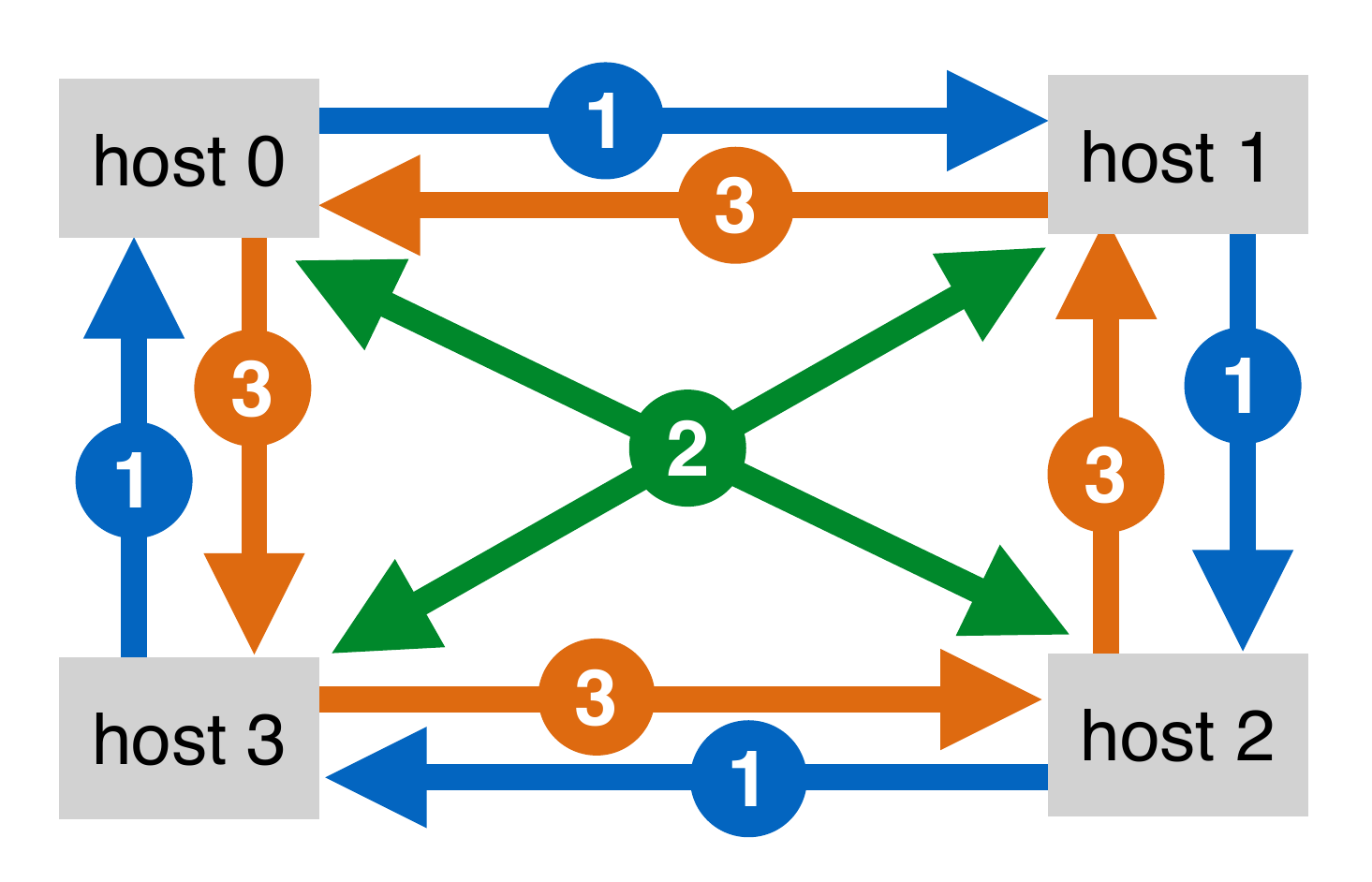}
	\label{Figure:RoundRobin}
}
\hfil
\subfigure[Application-level network scheduling improves throughput by up to 40\,\%] {
\centering
\begin{tikzpicture}
\begin{axis}[
	height=4cm,
	width=5.5cm,
	axis x line*=bottom,
	axis y line*=left,
	xlabel={number of servers},
	ylabel={throughput in GB/s},
	xmin=2,
	xmax=8,
	ymin=0,
	ymax=4,
	ymajorgrids,
	xtick={2, 3, 4, 5, 6, 7, 8},
	tick align=center,
	legend style={anchor=north west, draw=none, fill=white, at={(0, .24)}},
	legend cell align=left,
	legend columns=-1
]
\addplot[ultra thick, mark=*, color3, /pgfplots/error bars/.cd, x dir=none, y dir=both, y explicit, error bar style={line width=1pt}, error mark options={rotate=90, mark size=4pt, line width=1pt}] table [x index={0}, y index={3}, y error index={4}] {plots/scheduling.txt};
\addlegendentry{all-to-all}
\addplot[ultra thick, mark=square*, color1] table [x index={0}, y index={1}, y error index={2}] {plots/scheduling.txt};
\addlegendentry{round-robin}
\end{axis}
\end{tikzpicture}
\label{Figure:NetworkScheduling}
}
\hfil
\subfigure[512\,KB messages or larger hide synchronization cost completely (6 servers)] {
\centering
\begin{tikzpicture}
\begin{axis}[
	height=4cm,
	width=5.5cm,
	axis x line*=bottom,
	axis y line*=left,
	xlabel={message size},
	ylabel={throughput in GB/s},
	xmin=10,
	xmax=26,
	xtick={10, 14, 18, 22, 26},
	xticklabels={{1\,KB}, {16\,KB}, {256\,KB}, {4\,MB}, {64\,MB}},
	ymin=0,
	ymax=4,
	ymajorgrids,
	tick align=center
]
\addplot[ultra thick, mark=square*, color1] table [x index={0}, y index={1}, y error index={2}] {plots/scheduling-size.txt};
\addplot[nodes near coords=512\,KB, every node near coord/.append style={fill=none, anchor=270, inner sep=.13cm}, ultra thick, mark=square*, color2] coordinates {(19, 3.3362)};
\end{axis}
\end{tikzpicture}
\label{Figure:MessagesPerPhase}
}
\caption{Application-level network scheduling avoids switch contention, improving throughput by 40\,\%}
\end{figure*}
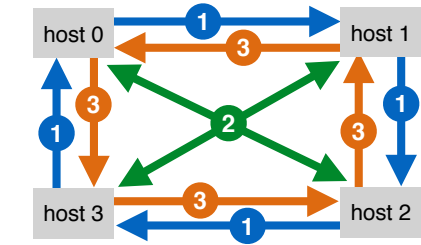
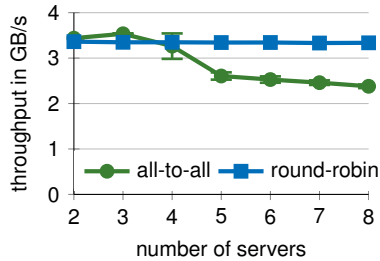
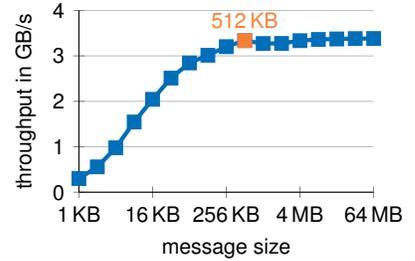

Modern servers with large main-memory capacities feature a non-uniform memory architecture (NUMA). Every CPU has its own local memory controller and accesses remote memory via QPI links that connect CPUs. As QPI speed is slower than local memory and has a higher latency, a remote access is more expensive than a local access. The query execution engine has to take this into account and restrict itself to local memory accesses as much as possible. Our communication multiplexer exposes NUMA characteristics to the database system to avoid incurring this performance penalty. The multiplexer has one receive queue for every NUMA socket as shown in Figure~\ref{Figure:HybridParallelism} and alternatively receives messages for each of them. This also means that NUMA is hidden inside the server so that servers in the cluster could potentially have heterogenous architectures. The multiplexer supports work-stealing: workers take messages from remote queues when their NUMA-local queue is empty.

For our 6-server cluster, allocating messages on a single socket reduces TPC-H performance for the hash-join plans by a mere 8\,\%. However, these servers have only two sockets that are further well-connected via two QPI links. This explains the minimal NUMA effects. Figure~\ref{Figure:NUMA} shows the measurement for a 4-socket Sandy Bridge EP server with 15 cores per socket and 1\,TB of main memory. Sockets are fully-connected with one QPI link for each combination of sockets. Interleaved allocation of the network buffers reduces TPC-H performance by 17\,\% compared to NUMA-aware allocation, allocating messages on a single socket even by 52\,\%. This demonstrates NUMA-aware allocation of message buffers can have a huge impact on performance. Our novel communication multiplexer therefore provides NUMA-local message buffers to the decoupled exchange operators.


\subsubsection{Application-Level Network Scheduling}
\label{Section:NetworkScheduling}

Uncoordinated all-to-all network traffic can cause switch contention and reduce throughput significantly---even for non-blocking switches that have enough capacity to support all ports simultaneously at maximum throughput. In the case of Ethernet switches, input queuing in the switch can cause head-of-line (HOL) blocking. InfiniBand implements a credit-based link-level flow control and no data is transmitted unless the available credits indicate sufficient buffer space at the receiver. While this prevents head-of-line blocking, switch contention is still possible: When several input ports transmit data to the same output port, the credits from the corresponding receiver run out faster than they are granted. Other packets from the same input ports could still be processed, however, the buffer space for input ports is limited. Thus, it is possible that all outstanding packets of a port run out of credits. This creates back pressure and the switch cannot receive more packets for this input port until it obtains new credits from the receiver.

Network scheduling with global knowledge of all active flows has been proposed before to solve the problem of switch contention. Hedera~\cite{AlFares2013Hedera} uses a central coordinator that regularly collects flow statistics and moves data flows from congested to underutilized links. However, flow estimation and scheduling is performed only every 5~seconds---much too infrequent for high-speed networks where transfers take a few milliseconds and a complete TPC-H run finishes in less than 5~seconds at scale factor 100. Neo-Join~\cite{Roediger2014NeoJoin} uses application-level network scheduling solving the Open Shop problem to minimize join execution time. However, its scheduling algorithm requires prior knowledge of data transfer sizes and does not scale well as its runtime is in $\mathcal{O}(n^4)$ for $n$ servers.

High-speed networks require a new approach to network scheduling that reacts fast and incurs latencies of at most a few microseconds similar to NUMA shuffling inside a single server \cite{Li2013NUMAShuffling}. For this reason we decided to implement a simple but very efficient round-robin network scheduling algorithm that makes use of special low-latency RDMA operations. It avoids HOL blocking for Ethernet and credit starvation for InfiniBand by dividing communication into distinct phases that prevent link sharing. In each phase a server has one target to which it sends, and a single source from which it receives as shown in Figure~\ref{Figure:RoundRobin} for four servers and three phases. Round-robin scheduling improves throughput by up to 40\,\% for an 8-server InfiniBand 4$\times$QDR cluster as demonstrated by the micro-benchmark in Figure~\ref{Figure:NetworkScheduling}. In this experiment, we added two smaller servers to our cluster to fully utilize our 8-port InfiniBand switch. Each server transmits 1,680 messages of size 512\,KB. After sending 8 messages to a fixed target, all servers synchronize via low-latency ($\sim$1\,$\mu s$) inline synchronization messages before they send to the next target. The data transfer between synchronizations has to be large enough to amortize the time needed for synchronization as illustrated in Figure~\ref{Figure:MessagesPerPhase}. For our distributed engine we thus use a message size of 512\,KB. It is important to reduce the CPU overhead of handling completion notifications for synchronization messages by processing only every $n$th completion notification. We use the maximum of 16k active work requests supported by our hardware to keep the synchronization latency at a few microseconds.


\section{Evaluation}
\label{Section:Evaluation}

We integrated our distributed query processing engine in HyPer, a full-fledged main-memory database system that supports the SQL-92 standard and offers excellent single-server performance. The experiments focus on ad-hoc analytical distributed query processing performance.


\subsection{Experimental Setup}

We conducted all experiments on a cluster of six identical servers connected via ConnectX-3 host channel adapters (HCAs) to an 8-port QSFP InfiniScale IV InfiniBand IS5022 switch operating at 4$\times$ quad data rate (QDR) resulting in a theoretical network bandwidth of 4\,GB/s per link. Each Linux server (Ubuntu 14.10, kernel 3.16.0-41) is equipped with two Intel Xeon E5-2660 v2 CPUs clocked at 2.20 GHz with 10 physical cores (20 hardware contexts due to hyper-threading) and 256 GB of main memory---resulting in a total of 120 cores (240 hardware contexts) and 1.5 TB of main memory in the cluster. The hardware setup is illustrated in Figure~\ref{Figure:Setup}. The thickness of a line in the diagram corresponds to the respective bandwidth of the connection.

HyPer offers both row and column-wise data storage; all experiments were conducted using the columnar format and only primary key indexes were created. On each server, HyPer transparently distributes the input relations over all available NUMA sockets (two in our case). Execution times include memory allocation (from the OS), page faulting, and deallocation for intermediate results, hash tables, etc.

TPC-H joins relations mostly along key/foreign-key relationships and thus benefits considerably when relations are partitioned accordingly. This enables partially or even completely local joins that avoid network traffic and thus improve query response times. Still, we decided against partitioning relations for HyPer as this is a manual process that requires prior knowledge of the workload. Instead, we assign relation chunks to servers as generated by \emph{dbgen} without initial redistribution for all experiments in this paper.


\subsection{Hybrid Parallelism}

This section evaluates the performance of our new approach by analyzing the scalability of the individual TPC-H queries for different distributed query execution engines. We further analyze the impact of network scheduling.

\begin{figure*}[t]
\centering
\pgfplotstableread{plots/scalability.txt}{\scalability}
\begin{tikzpicture}
\begin{groupplot}[
	group style = {
		group size=8 by 3,
		xlabels at=edge bottom,
	        xticklabels at=edge bottom,
		ylabels at=edge left,
	        yticklabels at=edge left,
	        horizontal sep=.2cm,
	        vertical sep=.2cm
	},
	axis x line*=bottom,
	axis y line*=left,
	title style={at={(.2, .62)}, fill=white},
	height=3.5cm,
	width=3.5cm,
	xmin=1,
	xmax=6,
	xtick={2, 4, 6},
	extra x ticks={1, 3, 5},
	extra x tick labels={},
	ymin=0,
	ymax=6,
	ymajorgrids,
	ytick={1, 2, 3, 4, 5, 6},
	tick align=center
]
\nextgroupplot[title=Q1]
	\addplot[ultra thick, mark=square*, color1] table [x={nodes}, y={q1-rdma}] {\scalability};
	\addplot[ultra thick, mark=*, color3] table [x={nodes}, y={q1-ipoib}] {\scalability};
	\addplot[ultra thick, mark=triangle*, color2] table [x={nodes}, y={q1-tcp}] {\scalability};
\nextgroupplot[title=Q2]
	\addplot[ultra thick, mark=square*, color1] table [x={nodes}, y={q2-rdma}] {\scalability};
	\addplot[ultra thick, mark=*, color3] table [x={nodes}, y={q2-ipoib}] {\scalability};
	\addplot[ultra thick, mark=triangle*, color2] table [x={nodes}, y={q2-tcp}] {\scalability};
\nextgroupplot[title=Q3]
	\addplot[ultra thick, mark=square*, color1] table [x={nodes}, y={q3-rdma}] {\scalability};
	\addplot[ultra thick, mark=*, color3] table [x={nodes}, y={q3-ipoib}] {\scalability};
	\addplot[ultra thick, mark=triangle*, color2] table [x={nodes}, y={q3-tcp}] {\scalability};
\nextgroupplot[title=Q4]
	\addplot[ultra thick, mark=square*, color1] table [x={nodes}, y={q4-rdma}] {\scalability};
	\addplot[ultra thick, mark=*, color3] table [x={nodes}, y={q4-ipoib}] {\scalability};
	\addplot[ultra thick, mark=triangle*, color2] table [x={nodes}, y={q4-tcp}] {\scalability};
\nextgroupplot[title=Q5]
	\addplot[ultra thick, mark=square*, color1] table [x={nodes}, y={q5-rdma}] {\scalability};
	\addplot[ultra thick, mark=*, color3] table [x={nodes}, y={q5-ipoib}] {\scalability};
	\addplot[ultra thick, mark=triangle*, color2] table [x={nodes}, y={q5-tcp}] {\scalability};
\nextgroupplot[title=Q6]
	\addplot[ultra thick, mark=square*, color1] table [x={nodes}, y={q6-rdma}] {\scalability};
	\addplot[ultra thick, mark=*, color3] table [x={nodes}, y={q6-ipoib}] {\scalability};
	\addplot[ultra thick, mark=triangle*, color2] table [x={nodes}, y={q6-tcp}] {\scalability};
\nextgroupplot[title=Q7]
	\addplot[ultra thick, mark=square*, color1] table [x={nodes}, y={q7-rdma}] {\scalability};
	\addplot[ultra thick, mark=*, color3] table [x={nodes}, y={q7-ipoib}] {\scalability};
	\addplot[ultra thick, mark=triangle*, color2] table [x={nodes}, y={q7-tcp}] {\scalability};
\nextgroupplot[title=Q8]
	\addplot[ultra thick, mark=square*, color1] table [x={nodes}, y={q8-rdma}] {\scalability};
	\addplot[ultra thick, mark=*, color3] table [x={nodes}, y={q8-ipoib}] {\scalability};
	\addplot[ultra thick, mark=triangle*, color2] table [x={nodes}, y={q8-tcp}] {\scalability};
\nextgroupplot[title=Q9, ylabel={speed-up of query response times}]
	\addplot[ultra thick, mark=square*, color1] table [x={nodes}, y={q9-rdma}] {\scalability};
	\addplot[ultra thick, mark=*, color3] table [x={nodes}, y={q9-ipoib}] {\scalability};
	\addplot[ultra thick, mark=triangle*, color2] table [x={nodes}, y={q9-tcp}] {\scalability};
\nextgroupplot[title=Q10]
	\addplot[ultra thick, mark=square*, color1] table [x={nodes}, y={q10-rdma}] {\scalability};
	\addplot[ultra thick, mark=*, color3] table [x={nodes}, y={q10-ipoib}] {\scalability};
	\addplot[ultra thick, mark=triangle*, color2] table [x={nodes}, y={q10-tcp}] {\scalability};
\nextgroupplot[title=Q11]
	\addplot[ultra thick, mark=square*, color1] table [x={nodes}, y={q11-rdma}] {\scalability};
	\addplot[ultra thick, mark=*, color3] table [x={nodes}, y={q11-ipoib}] {\scalability};
	\addplot[ultra thick, mark=triangle*, color2] table [x={nodes}, y={q11-tcp}] {\scalability};
\nextgroupplot[title=Q12]
	\addplot[ultra thick, mark=square*, color1] table [x={nodes}, y={q12-rdma}] {\scalability};
	\addplot[ultra thick, mark=*, color3] table [x={nodes}, y={q12-ipoib}] {\scalability};
	\addplot[ultra thick, mark=triangle*, color2] table [x={nodes}, y={q12-tcp}] {\scalability};
\nextgroupplot[title=Q13]
	\addplot[ultra thick, mark=square*, color1] table [x={nodes}, y={q13-rdma}] {\scalability};
	\addplot[ultra thick, mark=*, color3] table [x={nodes}, y={q13-ipoib}] {\scalability};
	\addplot[ultra thick, mark=triangle*, color2] table [x={nodes}, y={q13-tcp}] {\scalability};
\nextgroupplot[title=Q14]
	\addplot[ultra thick, mark=square*, color1] table [x={nodes}, y={q14-rdma}] {\scalability};
	\addplot[ultra thick, mark=*, color3] table [x={nodes}, y={q14-ipoib}] {\scalability};
	\addplot[ultra thick, mark=triangle*, color2] table [x={nodes}, y={q14-tcp}] {\scalability};
\nextgroupplot[title=Q15]
	\addplot[ultra thick, mark=square*, color1] table [x={nodes}, y={q15-rdma}] {\scalability};
	\addplot[ultra thick, mark=*, color3] table [x={nodes}, y={q15-ipoib}] {\scalability};
	\addplot[ultra thick, mark=triangle*, color2] table [x={nodes}, y={q15-tcp}] {\scalability};
\nextgroupplot[title=Q16]
	\addplot[ultra thick, mark=square*, color1] table [x={nodes}, y={q16-rdma}] {\scalability};
	\addplot[ultra thick, mark=*, color3] table [x={nodes}, y={q16-ipoib}] {\scalability};
	\addplot[ultra thick, mark=triangle*, color2] table [x={nodes}, y={q16-tcp}] {\scalability};
\nextgroupplot[title=Q17]
	\addplot[ultra thick, mark=square*, color1] table [x={nodes}, y={q17-rdma}] {\scalability};
	\addplot[ultra thick, mark=*, color3] table [x={nodes}, y={q17-ipoib}] {\scalability};
	\addplot[ultra thick, mark=triangle*, color2] table [x={nodes}, y={q17-tcp}] {\scalability};
\nextgroupplot[title=Q18]
	\addplot[ultra thick, mark=square*, color1] table [x={nodes}, y={q18-rdma}] {\scalability};
	\addplot[ultra thick, mark=*, color3] table [x={nodes}, y={q18-ipoib}] {\scalability};
	\addplot[ultra thick, mark=triangle*, color2] table [x={nodes}, y={q18-tcp}] {\scalability};
\nextgroupplot[title=Q19]
	\addplot[ultra thick, mark=square*, color1] table [x={nodes}, y={q19-rdma}] {\scalability};
	\addplot[ultra thick, mark=*, color3] table [x={nodes}, y={q19-ipoib}] {\scalability};
	\addplot[ultra thick, mark=triangle*, color2] table [x={nodes}, y={q19-tcp}] {\scalability};
\nextgroupplot[title=Q20, xlabel={number of servers}]
	\addplot[ultra thick, mark=square*, color1] table [x={nodes}, y={q20-rdma}] {\scalability};
	\addplot[ultra thick, mark=*, color3] table [x={nodes}, y={q20-ipoib}] {\scalability};
	\addplot[ultra thick, mark=triangle*, color2] table [x={nodes}, y={q20-tcp}] {\scalability};
\nextgroupplot[title=Q21]
	\addplot[ultra thick, mark=square*, color1] table [x={nodes}, y={q21-rdma}] {\scalability};
	\addplot[ultra thick, mark=*, color3] table [x={nodes}, y={q21-ipoib}] {\scalability};
	\addplot[ultra thick, mark=triangle*, color2] table [x={nodes}, y={q21-tcp}] {\scalability};
\nextgroupplot[title=Q22, legend style={align=left, anchor=north west, draw=none, fill=white, at={(1.05, .9)}}, legend cell align=left, legend columns=1]
	\addplot[ultra thick, mark=square*, color1] table [x={nodes}, y={q22-rdma}] {\scalability};
	\addlegendentry{RDMA (40 Gb/s InfiniBand)\\+ network scheduling}
	\addplot[ultra thick, mark=*, color3] table [x={nodes}, y={q22-ipoib}] {\scalability};
	\addlegendentry{TCP/IP (40 Gb/s InfiniBand)}
	\addplot[ultra thick, mark=triangle*, color2] table [x={nodes}, y={q22-tcp}] {\scalability};
	\addlegendentry{TCP/IP (1 Gb/s Ethernet)}
\end{groupplot}
\end{tikzpicture}
\caption{Scalability of the individual TPC-H queries for different query execution engines (HyPer, SF 100)}
\label{Figure:TPCHScalability}
\end{figure*}
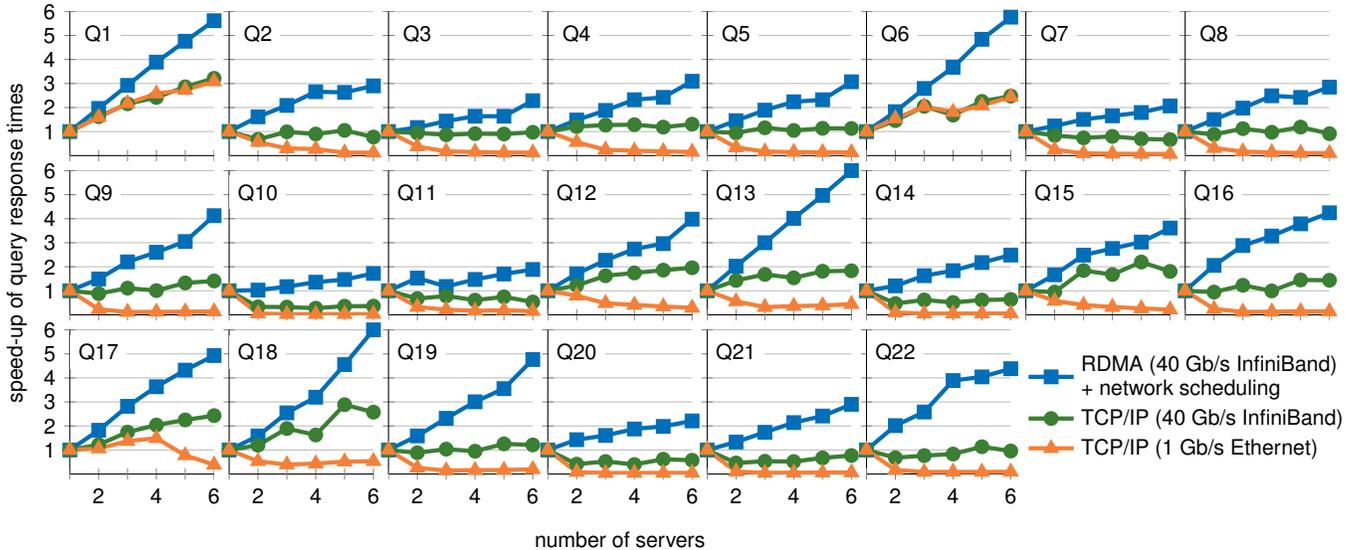

\subsubsection{Scalability}

Figure~\ref{Figure:TPCHScalability} shows how the individual TPC-H queries scale when servers are added to the cluster. It is apparent that the queries do not scale for Gigabit Ethernet. The only exceptions are Q1 and Q6, which transfer almost no data over the network. The scalability graph of Q17 illustrates the impact of switch contention: For more than 4 servers the effective bandwidth drops significantly and thus also the query performance. TCP/IP over InfiniBand (4$\times$QDR) performs better than TCP over Gigabit Ethernet but still does not scale well, mostly staying close to single-server performance or even performing worse (e.g, for Q10, Q11, and Q20). Only our RDMA-based communication multiplexer with network scheduling can improve the performance for all queries, with an overall speed-up of 3.5$\times$ for 6 servers (cf., Figure~\ref{Figure:TPCHSpeedUp}).

\subsubsection{Network Scheduling}

We analyzed the impact of network scheduling on HyPer's TPC-H performance for our 6-server cluster. Scheduling improves HyPer's TPC-H performance by 230\,\% when Gigabit Ethernet is used to connect the servers. Due to the high CPU overhead of TCP stack processing for TCP/IP over InfiniBand, network scheduling does not improve performance in this setting. For RDMA, network scheduling improves HyPer's TPC-H performance by 12.2\,\%. We expect the impact of network scheduling on TPC-H performance to increase further with the cluster size.


\subsection{Distributed SQL Systems}

We compare HyPer with four state-of-the-art distributed SQL systems: Spark SQL 1.3, Cloudera Impala 2.2 \cite{Kornacker2015Impala}, MemSQL 4.0, and Vectorwise Vortex \cite{Costea2012VectorwiseMPP}. HyPer, MemSQL, and Vectorwise use custom data storage. We ensure that Spark SQL caches the HDFS input as deserialized Java objects in main memory before query execution (cache level \texttt{MEMORY\_ONLY}) to avoid deserialization overheads. Impala processes HDFS-resident Parquet files during query execution. We ensured a hot Linux buffer cache to avoid expensive disk accesses. Still, Impala has to perform deserialization during query execution. We conducted a micro-benchmark analyzing multiple TPC-H queries and found that deserialization makes up less than 30\,\% of the query execution time.

Spark SQL and Impala are installed as part of the Cloudera Hadoop distribution (CDH 5.4). We set the HDFS replication factor to 3 and enabled short-circuit reads. The cluster is configured so that all systems use the high-speed InfiniBand interconnect instead of Gigabit Ethernet. We use unmodified TPC-H queries except for Spark SQL---which in three cases required rewritten TPC-H queries that avoid correlated subqueries \cite{Floratou2014SQLHadoop}---and a modified query 11 for Impala, which does not support subqueries in the having clause.

\begin{figure*}
\centering
\subfigure[Queries per hour for each distributed SQL system] {
\begin{tikzpicture}
\begin{axis}[
	xbar=0cm,
	axis x line=bottom,
	axis y line*=left,
	bar width=.2cm,
	symbolic y coords={SparkSQL, Impala, MemSQL, Vectorwise, Hyper1, Hyper2},
	yticklabels={Spark SQL, Impala, MemSQL, Vectorwise, HyPer (chunked), HyPer (partitioned)},
	width=7.5cm,
	height=4.5cm,
	enlarge y limits=.1,
	xlabel={queries per hour},
	yticklabel style={align=center},
	ytick=data,
	xmin=0,
	xmax=26,
	xtick={0, 5, 10, 15, 20, 25},
	xticklabels={0, 5k, 10k, 15k, 20k, 25k},
	xmajorgrids,
	tick align=center
]
\addplot[fill=color1, draw=color1] coordinates {
(0.077,SparkSQL)
(0.123,Impala)
(0.544,MemSQL)
(3.856,Vectorwise)
(16.090,Hyper1)
(20.739,Hyper2)
};
\node[right,fill=white] at (axis cs:.2,SparkSQL) {77};
\node[right,fill=white] at (axis cs:.2,Impala) {123};
\node[right,fill=white] at (axis cs:.6,MemSQL) {544};
\node[right,fill=white] at (axis cs:4,Vectorwise) {3,856};
\node[right,fill=white] at (axis cs:16.1,Hyper1) {16,090};
\node[right,fill=white] at (axis cs:21,Hyper2) {20,739};
\end{axis}
\end{tikzpicture}
\label{Figure:CompetitorsTPCHQueries}
}
\hfil
\subfigure[Impact of network bandwidth on TPC-H performance] {
\centering
\begin{tikzpicture}
\begin{axis}[
	height=4.2cm,
	width=8cm,
	axis x line*=bottom,
	axis y line=left,
	xlabel={network bandwidth},
	ylabel={speed-up over GbE},
	xmin=0,
	xmax=4,
	ymin=0,
	ymax=13,
	ymajorgrids,
	ytick={1, 5, 10},
	yticklabels={{1$\times$}, {5$\times$}, {10$\times$}},
	xtick={0.125, 1, 2, 4},
	xticklabels={{GbE\\0.125 GB/s}, {SDR\\1 GB/s}, {DDR\\2 GB/s}, {QDR\\4 GB/s}},
	xticklabel style={align=center},
	tick align=center,
	legend style={anchor=north west, draw=none, fill=none, at={(.03, 1.05)}, inner sep=0pt},
	legend cell align=left,
	legend columns=2
]
\addplot[ultra thick, mark=square*, color1] coordinates {
	(0.125, 1)
	(1, 4.66)
	(2, 7.33)
	(4, 12.19)
};
\addlegendentry{HyPer (RDMA)}
\addplot[ultra thick, mark=*, color3] coordinates {
	(0.125, 1)
	(1, 3.21)
	(2, 3.75)
	(4, 3.79)
};
\addlegendentry{HyPer (TCP)}
\addplot[ultra thick, mark=triangle*, color2] coordinates {
	(0.125, 1)
	(1, 3.46)
	(2, 3.58)
	(4, 3.61)
};
\addlegendentry{Vectorwise}
\addplot[ultra thick, mark=diamond*, color4] coordinates {
	(0.125, 1)
	(1, 1.13)
	(2, 1.14)
	(4, 1.23)
};
\addlegendentry{MemSQL}
\end{axis}
\end{tikzpicture}
\label{Figure:TPCHNetworkBandwidth}
}
\caption{Comparing distributed analytical SQL systems for the TPC-H benchmark (6 servers, SF 100)}
\label{Figure:CompetitorsTPCH}
\end{figure*}
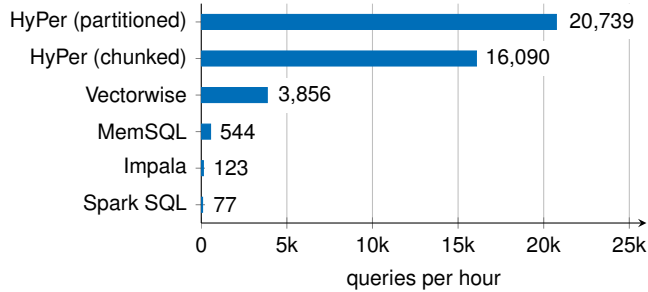
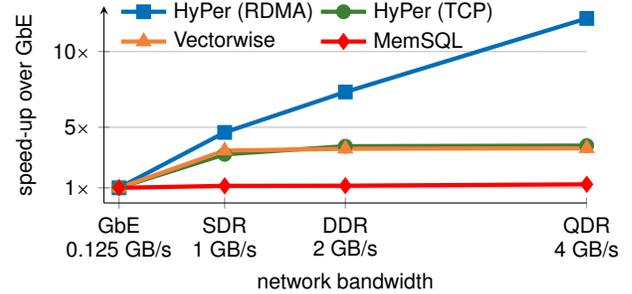

\begin{table*}[t!]
\centering
\newcolumntype{Y}{>{\raggedleft\arraybackslash}X}
\begin{tabularx}{\linewidth}{lYYYYYY}
\toprule
 & {Spark SQL} & {Impala} &  {MemSQL} & {Vectorwise} & {HyPer} & {HyPer}\\
\midrule
data placement & chunked & chunked & partitioned & partitioned & chunked & partitioned\\
\midrule
Q1 & 7.09s & 7.04s & 8.38s & 0.80s & \textbf{0.08s} & 0.10s\\
Q2 & 16.92s & 8.91s & 1.65s & 0.37s & \textbf{0.04s} & 0.05s\\
Q3 & 26.04s & 23.79s & 13.90s & 0.24s & 0.34s & \textbf{0.18s}\\
Q4 & 10.21s & 25.53s & 0.81s & \textbf{0.06s} & 0.16s & 0.18s\\
Q5 & 1m 13.17s & 23.57s & 8.83s & 0.84s & 0.22s & \textbf{0.13s}\\
Q6 & 1.97s & 3.39s & 2.50s & 0.05s & \textbf{0.03s} & 0.05s\\
Q7 & 59.67s & 43.57s & 2.76s & 0.36s & 0.33s & \textbf{0.16s}\\
Q8 & 1m 20.01s & 21.04s & 2.48s & 2.01s & 0.17s & \textbf{0.10s}\\
Q9 & 2m 52.28s & 1m 11.12s & 11.92s & 1.44s & 0.60s & \textbf{0.56s}\\
Q10 & 18.63s & 8.86s & 1.50s & 1.56s & 0.51s & \textbf{0.26s}\\
Q11 & -- & 4.20s & 0.53s & 0.22s & \textbf{0.06s} & 0.12s\\
Q12 & 18.30s & 8.97s & 1.76s & 0.10s & 0.12s & \textbf{0.09s}\\
Q13 & 12.36s & 25.97s & 4.48s & 3.61s & \textbf{0.36s} & 0.41s\\
Q14 & 7.13s & 6.00s & 2.29s & 0.69s & 0.06s & \textbf{0.05s}\\
Q15 & 12.92s & 4.83s & 13.00s & 0.95s & \textbf{0.08s} & \textbf{0.08s}\\
Q16 & 11.37s & 7.35s & 3.44s & 0.69s & 0.20s & \textbf{0.16s}\\
Q17 & 2m 20.00s & 1m 27.96s & 0.75s & 0.53s & \textbf{0.09s} & 0.11s\\
Q18 & 1m 39.66s & 1m 06.79s & 51.30s & 1.63s & 0.58s & \textbf{0.37s}\\
Q19 & 8.73s & 1m 43.13s & 0.60s & 0.81s & \textbf{0.20s} & 0.23s\\
Q20 & 24.97s & 16.27s & 8.53s & 0.51s & 0.14s & \textbf{0.13s}\\
Q21 & 2m 49.93s & 1m 07.59s & 2.08s & 1.91s & 0.49s & \textbf{0.24s}\\
Q22 & 15.69s & 6.35s & 2.19s & 1.18s & 0.07s & \textbf{0.06s}\\
\midrule
packets sent & 106.58 million & 175.65 million & 7.02 million & 7.06 million & 7.3 million & 2.4 million\\
data shuffled & 211.30\,GB & 140.52\,GB & 13.96\,GB & 19.36\,GB & 27.95\,GB & 8.88\,GB\\
disk I/O & 0.23\,GB & 0.04\,GB & 0.03\,GB & 0.01\,GB & 0.00\,GB & 0.00\,GB\\
\midrule
total time & 16m\,19s & 10m\,42s & 2m\,26s & 20.54s & 4.92s & \textbf{3.82s}\\
geometric mean & 24.32 & 17.21 & 3.23 & 0.59 & 0.16 & \textbf{0.14}\\
queries per hour & 77 & 123 & 544 & 3,856 & 16,090 & \textbf{20,739}\\
\bottomrule
\end{tabularx}\par\medskip
\caption{Detailed query runtimes for a complete TPC-H run; fastest runtime in bold (6 servers, SF 100)}
\label{Table:CompetitorsTPCHQueries}
\end{table*}

Figure~\ref{Figure:CompetitorsTPCHQueries} compares the five systems for the TPC-H benchmark on a scale factor 100 data set ($\sim$110\,GB data on disk). A full TPC-H run takes 16~min 19~sec for Spark SQL (without query 11), 10~min 42~sec for Impala, 2~min 26~sec for MemSQL, 20.5~sec for Vectorwise, 4.9~sec for HyPer with chunked data placement and 3.8~sec when relations are partitioned by the first attribute of the primary key. Table~\ref{Table:CompetitorsTPCHQueries} shows detailed query execution times with the fastest in bold as well as network and disk I/O metrics. Queries are run 10 times, keeping the median. The Linux file system cache is not flushed between runs to keep data hot in main memory.

\subsubsection{Configuration Details}

We tuned all compared systems according to their documentation, previous publications, and direct advice from their developers to ensure their best possible performance.

\textbf{Spark SQL.} Apache Spark SQL, the follow-up to Shark, is still early in development. We ensure in-memory processing by specifying the tables as temporary and explicitly caching them. We further configured Spark to use main memory for its scratch space and disabled spilling to disk. We increased the per executor memory to 250 GB per server and the parallelism level to twice the number of cores in the cluster as recommended. For queries 2, 15, and 22 we had to use rewritten queries from \cite{Floratou2014SQLHadoop} that avoid correlated subqueries. Spark SQL takes more than an hour for query 11, we thus measured all metrics for the remaining 21 queries.

\textbf{Impala.} Impala is run with generated statistics on Parquet files with disabled pretty printing. Short-circuit reads and runtime code generation are enabled, operator spilling disabled. Using the HDFS cache did not improve query performance compared to the standard Linux file system cache.

\textbf{MemSQL.} We configured MemSQL to replicate nation and region as reference tables. All other relations are partitioned by the first column of their primary key. We used one partition per CPU core in the cluster as recommended in the MemSQL documentation for maximum parallelism. We further created foreign key indexes to enable index-nested-loop joins that improve performance significantly.

\textbf{Vectorwise.} We measure Vectorwise Vortex with statistics, primary keys, and foreign keys. We set the maximum parallelism level to 120 and the number of cores to 20 as recommended by an Actian engineer for our cluster. Similar to MemSQL, all relations except nation and region are partitioned by the first column of their primary key. Replicating customer and supplier as recommended by Actian for an optimal TPC-H performance reduces the runtime by 19\,\% to 16.56 seconds and the geometric mean to 0.5.

\textbf{HyPer.} For HyPer we measured both chunked placement to compare against Spark SQL and Impala as well as partitioned placement to compare against MemSQL and Vectorwise. For chunked relations, HyPer has to use distributed joins and aggregations, shuffling considerably more data than Vectorwise and MemSQL (see Table~\ref{Table:CompetitorsTPCHQueries}) but still outperforms them due to its fast RDMA-based query engine.


\subsubsection{Network Bandwidth}

Our InfiniBand hardware supports data rates of 1\,GB/s (single data rate), 2\,GB/s (double data rate), and 4\,GB/s (quad data rate). Figure~\ref{Figure:TPCHNetworkBandwidth} shows the impact of the network bandwidth on the TPC-H performance of the in-memory MPP database systems HyPer, Vectorwise Vortex, and MemSQL. For each system, we show the speed up compared to its performance using Gigabit Ethernet. MemSQL only improves by 23\,\% when using the 32$\times$ faster InfiniBand 4$\times$QDR interconnect. Vectorwise Vortex and a variant of HyPer using TCP achieve a speed up of 4$\times$ for InfiniBand but cannot scale performance substantially when increasing the data rate. Our RDMA-enabled communication multiplexer enables HyPer to scale its TPC-H performance with the network bandwidth, processing 12$\times$ more queries per hour for InfiniBand 4$\times$QDR compared to Gigabit Ethernet.


\subsubsection{Larger Scale Factor}
\label{Section:LargerScaleFactor}

We further ran TPC-H at scale factor 300 for the three fastest systems to see how they scale to larger inputs: HyPer takes 12~sec to process the $\sim$320\,GB of data. This is 3.1$\times$ more than the 3.8~sec for SF 100. Vectorwise Vortex takes 44.2~sec for SF 300, an increase of 2.2$\times$ compared to 20.5~sec for SF 100. MemSQL processes a SF 300 data set in 8~min 46~sec, this is 3.4$\times$ more than the 2~min 26~sec for SF 100.


\section{Related Work}
\label{Section:RelatedWork}

Parallel databases are a well-studied field of research that attracted considerable attention in the 1980s/90s with Grace \cite{Fushimi1986Grace}, Gamma \cite{DeWitt1990Gamma}, Bubba \cite{Boral1990Bubba}, Volcano \cite{Graefe1990Exchange}, and Prisma \cite{Apers1992PRISMA}. Today's commercial parallel database systems include Teradata, Exasol, IBM DB2, Oracle \cite{Mukherjee2015DistributedOracle}, Greenplum, SAP HANA \cite{Mukherjee2015DistributedOracle}, HP Vertica (which evolved from C-Store \cite{Stonebraker2005CStore}), MemSQL, Cloudera Impala \cite{Kornacker2015Impala}, and Vectorwise Vortex \cite{Costea2012VectorwiseMPP}.

The comparatively low bandwidth of standard network interconnects such as Gigabit Ethernet creates a bottleneck for distributed query processing. Consequently, recent research focused on minimizing network traffic: Neo-Join \cite{Roediger2014NeoJoin} and Track Join \cite{Polychroniou2014TrackJoin} decide during query processing how to redistribute tuples to exploit locality in the data placement. High-speed networks such as InfiniBand and RDMA remove this bottleneck and have been applied to database systems before. Frey et al.~\cite{Frey2010iWARP} designed the Cyclo Join for join processing within a ring topology. Goncalves and Kersten \cite{Goncalves2011DataCyclotron} extended MonetDB with a novel distributed query processing scheme based on continuously rotating data in a modern RDMA network with a ring topology. M\"uhl\-eisen et al.~\cite{Muehleisen2013RemoteMemory} pursued a different approach, using RDMA to utilize remote main memory for temporary database files in MonetDB. Kalia et al.~\cite{Kalia2014RDMAKeyValue} used RDMA to build a fast key-value store. Barthels et al.~\cite{Barthels2015RDMAJoin} provide a detailed analysis of a distributed radix join using RDMA for rack-scale InfiniBand clusters. Costea and Ionescu~\cite{Costea2012VectorwiseMPP} extended Vectorwise, which originated from the MonetDB/X100 project \cite{Zukowski2012Vectorwise}, to a distributed system using MPI over InfiniBand. The project is called Vectorwise Vortex and is included in our evaluation.

The problem of switch contention has been addressed in the literature before. Hedera \cite{AlFares2013Hedera} applies heuristics to move data flows from overloaded links to free links using a central coordinator with global knowledge. However, flow estimation and scheduling is performed only every 5~seconds---much too infrequent for high-speed networks where transfers take a few milliseconds and a complete TPC-H run finishes in less than 5~seconds at scale factor 100. Neo-Join~\cite{Roediger2014NeoJoin} uses application-level network scheduling solving the Open Shop problem to minimize join execution time. However, its scheduling algorithm requires prior knowledge of data transfer sizes and does not scale well as its runtime is in $\mathcal{O}(n^4)$ for $n$ servers. High-speed networks require the new approach to network scheduling described in this work that reacts fast and incurs latencies of at most a few microseconds.

Several papers discuss the implications of NUMA for data\-base systems. Albutiu et\,al.\ \cite{Albutiu2012MPSM} devised MPSM, a NUMA-aware sort-merge join algorithm. Li et\,al.\ \cite{Li2013NUMAShuffling} applied data shuffling to NUMA systems and particularly to MPSM. Our application-level network scheduling is similar to NUMA shuffling in that we also use a simple round-robin schedule to keep synchronization overheads at a few microseconds.

There has been various research analyzing TCP performance \cite{Clark1988TCPPerformance,Foong2003TCPPerformance,Frey2010iWARP}. We refer the reader to \cite{Frey2010iWARP} for a detailed discussion. Previous studies have found that TCP performance does not scale well to higher network bandwidths, as the receiver becomes CPU-bound. For large messages, data touching operations such as checksums and copying cause a high CPU load and the network interface card raises interrupts leading to many context switches \cite{Foong2003TCPPerformance,Frey2010iWARP}. TCP offloading, already proposed in the 1980s to alleviate the CPU bottleneck \cite{Clark1988TCPPerformance}, has since been implemented in hardware. A second problem identified in the literature is TCP's high memory bus load \cite{Clark1988TCPPerformance}: Every byte sent and received over the network causes 2-4 bytes traversing the memory bus \cite{Foong2003TCPPerformance}. However, our experiments have shown that the introduction of data direct I/O reduces memory bus traffic considerably for NUMA-aware applications (cf.~Section~\ref{Section:DDIO}).

IBM DB2 differentiates between local and global parallelism similar to hybrid parallelism to overcome some of the problems of the classic exchange operator. However, instead of decoupled exchange operators that enable work stealing, DB2 uses a special exchange operator that merges the results of threads to reduce the number of parallel units.


\section{Concluding Remarks}
\label{Section:Conclusion}

Remote direct memory access (RDMA) currently receives increasing interest in the database research community. Binnig et al.~\cite{Binnig2015EndOfSlowNetworks} have shown that database systems need to adopt RDMA to fully leverage the high bandwidth of InfiniBand. It is not enough to just use faster networking hardware, the software has to change as well to address the new bottlenecks that emerge. Recent work has shown the benefits of RDMA for specific relational operators (e.g., joins \cite{Barthels2015RDMAJoin}) and key-value stores \cite{Kalia2014RDMAKeyValue}. However, we are the first to present the design and implementation of a complete distributed query engine based on RDMA that is capable of processing complex analytical workloads such as the TPC-H benchmark. Our engine uses RDMA to avoid the overheads of TCP processing, low-latency network scheduling to address switch contention, and flexible parallelism to overcome the inflexibility of the classic exchange operator model. In combination, this allows us to scale the high single-server performance of a state-of-the-art in-memory database system with the number of servers in the cluster.

Oracle and SAP recently proposed approaches for hybrid distributed query and transaction processing, relying on a shared buffer cache \cite{Mukherjee2015DistributedOracle} and a shared log \cite{Goel2015DistributedHANA}, respectively. We plan to extend our work on hybrid processing \cite{Muehlbauer2013Scyper} from full replication to fragmented relations to scale query performance while sustaining HyPer's excellent TX throughput.


\section{Acknowledgments}

Wolf R\"odiger is a recipient of the Oracle External Research Fellowship. Tobias M\"uhlbauer is a recipient of the Google Europe Fellowship in Structured Data Analysis. This work has further been partially sponsored by the German Federal Ministry of Education and Research (BMBF) grant RTBI 01IS12057.

\balance


\bibliographystyle{abbrv-etal}
\bibliography{bibliography}

\end{document}